\documentclass[
 reprint,
superscriptaddress,
noeprint,
nobibnotes,
amsmath,amssymb,
prb,
]{revtex4-2}

\usepackage{graphicx}
\usepackage{dcolumn}
\usepackage{bm}
\usepackage{hyperref}
\usepackage{notes2bib}
\bibnotesetup{
  note-name = {},
  use-sort-key = false
}
\hypersetup{
    colorlinks=true,
    linkcolor=black,
    filecolor=black,      
    urlcolor=black,
    citecolor=black,
    pdftitle={Overleaf Example},
    }

\usepackage[dvipsnames]{xcolor}
\usepackage{notes2bib}

\begin{document}


\title{Excitation of confined bulk plasmons in metallic nanoparticles by penetrating electron beams within a nonlocal analytical approach}


\author{Mattin Urbieta}
\email[Contact author: ]{mattin.urbieta@ehu.eus}
 \affiliation{Department of Applied Physics, University of the Basque Country UPV/EHU,  Europa Plaza 1, 20018, Donostia-San Sebastián, Gipuzkoa, Spain.}

\author{Eduardo Ogando}%
\affiliation{%
 Department of Physics, University of the Basque Country UPV/EHU, Paseo de la Universidad 7, 01006, Vitoria-Gasteiz, Araba, Spain.
}%

\author{Alberto Rivacoba}
\affiliation{
 Donostia International Physics Center (DIPC), Paseo Manuel de Lardizabal 4, 20018, Donostia-San Sebastián, Gipuzkoa, Spain.
}%

\author{Javier Aizpurua}
 \email[Contact author: ]{aizpurua@ehu.eus}
 \affiliation{
 Donostia International Physics Center (DIPC), Paseo Manuel de Lardizabal 4, 20018, Donostia-San Sebastián, Gipuzkoa,  Spain.
}%
\affiliation{%
Department of Electricity and Electronics, FCT-ZTF, University of the Basque Country UPV/EHU, Barrio Sarriena z/g, 48940, Leioa, Bizkaia, Spain.
}%
 \affiliation{
 IKERBASQUE, Basque Foundation for Science, 48009, Bilbao, Bizkaia, Spain.
}%
\author{Nerea Zabala}
\email[Contact author: ]{nerea.zabala@ehu.eus}
\affiliation{
 Donostia International Physics Center (DIPC), Paseo Manuel de Lardizabal 4, 20018, Donostia-San Sebastián, Gipuzkoa, Spain.
}%
\affiliation{%
Department of Electricity and Electronics, FCT-ZTF, University of the Basque Country UPV/EHU, Barrio Sarriena z/g, 48940, Leioa, Bizkaia, Spain.
}%
\affiliation{
Materials Physics Center, CSIC-UPV/EHU, Manuel de Lardizabal 5, 20018, Donostia-San Sebastián, Gipuzkoa, Spain.
}%

\date{\today}

\begin{abstract}
Using a linear hydrodynamic model, we theoretically investigate the interaction between penetrating electron beams and sub-5-nm metallic spherical nanoparticles.
We derive an analytical expression of the electron energy loss probability that includes nonlocal effects in the response of the confined electron gas.
Our study focuses on the longitudinal plasmon excitations --also known as confined bulk plasmons (CBPs)-- which are inaccessible within local dielectric frameworks, and it shows that their excitation is highly sensitive to the impact parameter, the kinetic energy of the incident electron, and the nanoparticle size.
In contrast to local approaches, our model predicts the emergence of several peaks associated with CBPs above the plasma frequency and a blueshift of their spectral envelope as the nanoparticle size decreases and the impact parameter of the electron beam increases from its center toward the surface.
Moreover, it predicts a threshold impact parameter, corresponding to the minimum length of the electron's trajectory inside the nanoparticle required for efficient excitation of specific CBP modes.
By exploiting a multipolar description of the CBPs, we identify the underlying selection rules governing their excitation by electron beams and correlate the observed blueshift at larger impact parameters with the preferential excitation of higher-order CBPs.
The size-dependent dispersion of the CBPs further enhances this impact-parameter-dependent blueshift and also explains the decrease in the impact parameter threshold for smaller nanoparticles.
Compared with the local response approximation, the hydrodynamic model captures these trends without \textit{ad hoc} momentum cutoffs and with rapid multipolar convergence.
Applied to sodium nanoparticles as a canonical free‑electron system, the theory reproduces qualitative behaviors observed in experiments and \textit{ab initio} calculations, providing practical guidance for the design and interpretation of CBP‑sensitive electron energy loss spectroscopy measurements.
\end{abstract}

\maketitle


\section{\label{sec:level1}Introduction}

Our knowledge about the properties and dynamics of plasmons in nanoparticles is, to a large extent, due to advances in electron energy loss spectroscopy (EELS) \cite{Browning_1993_Nature, Batson_1993_Nature, Egerton_2011_book, Krivanek_2014_Nature} within scanning transmission electron microscopy (STEM) \cite{Batson_2002_Nature, Krivanek_2010_Nature} and in optical spectroscopies \cite{Grillet_2011_ACSNano, Sigle_2015_ACSNano}.
Over the last two decades, subnanometer spatial resolution \cite{Batson_2002_Nature, Nellist_2004_Science} and sub-eV energy sensitivity \cite{Hart_2017_SciRep} have been achieved, expanding the capabilities of STEM-EELS to unprecedented levels and opening new pathways to study novel materials and nanostructures \cite{Krivanek_2014_Nature, Ke_2015_BeilsteinJNanotechnol, Krivanek_2010_Nature, Rez_2016_NatComm, Hachtel_2019_Science, Lagos_2017_Nature, Govyadinov_2017_NatComm, Zeiger_2020_PhysRevLett, LourencoMartins_2017_PhysRevX}.
The study of localized surface plasmon (LSP) resonances in metallic nanoparticles \cite{Batson_1982_PhysRevLett, Batson_1982_Ultramicroscopy} has reached extraordinary spatial resolution, enabling EELS measurements on individual nanoparticles smaller than 10 nm \cite{Jiang_2009_SolidStateCommun, Scholl_2012_Nature, Raza_2013_Nanophotonics, Raza_2015_NatComm, Hobbs_2016_NanoLett, Campos_2019_NatPhys}, and establishing STEM-EELS as a powerful tool to explore complex phenomena at the nanometer scale.

Beyond LSPs, EELS can resolve confined bulk plasmons (CBPs) --longitudinal collective oscillations of the electron density confined to the volume within a nanoparticle-- also referred to as volume plasmons in the literature \cite{Hobbs_2016_NanoLett}.
These modes are typically inaccessible to optical spectroscopy techniques because their coupling to light is approximately 3 orders of magnitude weaker than that of dipolar LSPs \cite{Ruppin_1973_PhysRevLett,Christensen_2014_ACSNano}.
As a result, CBPs have received far less attention than their LSP counterparts.
In contrast, electron probes that penetrate the nanoparticle couple efficiently to CBPs and induce longitudinal perturbations of the electron cloud.
Experimentally, CBPs have been measured within STEM-EELS for a wide range of materials, including Ag \cite{Koh2009,Scholl_2012_Nature}, Al \cite{Batson_1985_SurfSci}, Bi \cite{Wang_2006_ApplPhysLett, BorjaUrby_2019_JElectronSpectrosRelatPhenomena}, and Si \cite{Mitome_1992_JApplPhys}.

From a theoretical perspective, the excitation of plasmons by penetrating electron beams in small nanoparticles has long been investigated within the local response approximation (LRA), both analytically \cite{Bausells_1987_SurfSci, Rivacoba_1990_ScannMicrosc, Stamatopoulou_2024_PhysRevRes} and numerically \cite{GarciadeAbajo_2002_PhysRevB, Hohenester_2014_CompPhysComm, Das_2012_JPhysChemC, Cao_2015_ACSPhotonics}.
In this framework, the material response is described by a local dielectric function $\varepsilon(\omega)$ that only depends on frequency $\omega$, typically modeled with the Drude form $\varepsilon(\omega) = \varepsilon_{\infty} - \omega_{p}^{2}/[\omega(\omega + i \gamma)]$, where $\omega_{p}$ is the bulk plasma frequency, $\gamma$ the damping term, and $\varepsilon_{\infty}$ the high-frequency background permittivity.
This approximation assumes an electron gas of infinite compressibility that reacts locally and independently to external perturbations, and it generally provides a reliable description of the photonic response for nanostructures larger than $\sim 10 \; \textrm{nm}$.
In the nonretarded limit, the energy loss experienced by a penetrating probe is obtained from the induced potential, which --under abrupt interface conditions-- follows from solving Poisson's equation with standard boundary conditions.
However, when the probe penetrates the target or travels very close to its surface, the LRA becomes inadequate: the local model cannot capture the reduced ability of conduction electrons to respond collectively to large wave‑vector components of the excitation field.
This deficiency produces an unphysical (logarithmic) divergence of the induced potential at the probe position, which is typically circumvented \textit{ad hoc} by imposing a momentum cutoff \cite{Pines_1952_PhysRev}.
In practice, bulk losses are modeled by adding a bulk plasmon term centered at the plasma energy of the electron gas and proportional to the electron path inside the nanoparticle \cite{GarciadeAbajo_2010_RevModPhys}.
In finite systems, the presence of a surface gives rise to the well-known Begrenzung effect, whereby the excitation probability of bulk plasmons is reduced due to the competing excitation of LSPs \cite{Ritchie_1957_PhysRev}.

As nanoparticles approach the sub-10-nm regime, the assumption of locality progressively breaks down and nonlocal effects --stemming from finite electron compressibility, electron diffusion, or spatial correlation-- become prominent.
In particular, CBPs arise from the finite compressibility of the electron gas: the associated internal pressure acts as a restoring force against rapid charge-density variations and drives longitudinal oscillations throughout the nanoparticle.
In the Thomas-Fermi description of the electron gas \cite{Raza_2011_PhysRevB, Raza_2015_JPhysCondensMatter}, this pressure sets a finite screening length that prevents electron-density accumulation at short length scales.
These modes lie outside the scope of the LRA and require nonlocal models that incorporate the wave-vector dependence of the electronic response, such as the linear hydrodynamic model\cite{Ruppin_1978_JPhysChemSolids}.
Consistent with this picture, the experimental observation of CBPs in the sub-5-nm regime is widely regarded as a hallmark of the hydrodynamic electron behavior \cite{Christensen_2014_ACSNano}, underscoring the need for deeper theoretical insight into their excitation by penetrating electron beams.

Beyond hydrodynamic descriptions, quantum \textit{ab initio} frameworks provide \cite{Urbieta_2024_PhysChemChemPhys, Candelas_2025_JPhysChemLett} a suitable platform to address the atomic-scale regime and the quantum nature of electrons and their influence on CBPs.
However, the interpretation of their results can be hindered by the complexity of the model, and the computational requirements are prohibitive for large systems. 
Recent studies based on \textit{ab initio} frameworks show that, while atomistic shape affects LSP features in the electron energy loss (EEL) spectra, the energies of CBPs are largely insensitive to particle shape or relative orientation for nanoparticles of the same size \cite{Urbieta_2024_PhysChemChemPhys}.
These studies also report that the energy of the bulk plasmon peak depends on the electron beam impact parameter --a behavior missed by LRA-based models-- but is instead well described by a linear hydrodynamic model.
Compared with approaches based on \textit{ab initio} frameworks, the hydrodynamic model offers analytical tractability and more straightforward data analysis, at the cost of applicability being limited to highly symmetric geometries.
Since the seminal work of Fujimoto and Komaki \cite{Fujimoto_1968_JPhysSocJpn}, who analyzed plasmon excitation in small spherical nanoparticles using broad electron beams, hydrodynamic treatments have been employed in several theoretical studies of the EEL spectra from nanoparticles excited by penetrating electron beams \cite{Barberan_1985_PhysRevB, TranThoai_1986_physstatussolidib, TranThoai_1988_physstatussolidia, V.Baltz_1995_ZPhysB, Gildenburg2016}.
Despite these advances, an analytical expression for the EEL probability of a spherical nanoparticle as a function of impact parameter for penetrating trajectories within a hydrodynamic model is still lacking.
Providing such an expression would enable systematic studies of CBPs and their excitation by electron beams within a transparent analytical framework.

In this work, we derive an analytical expression for the EEL probability of spherical metallic nanoparticles within a linear hydrodynamic model, valid for both penetrating and aloof electron beam trajectories.
We apply this expression to sodium (Na) nanoparticles and analyze the dependence of LSPs and, in particular, CBPs on the impact parameter, electron velocity, and nanoparticle size.
The nonlocal description introduced here captures essential features absent from local models,
the emergence of several peaks associated with CBPs above the plasma frequency, and a pronounced blueshift of their spectral envelope --bulk plasmon envelope-- with decreasing particle size and increasing impact parameter, as well as the existence of a minimum internal electron path length required to efficiently excite specific sets of CBPs.
Through a multipolar analysis, we identify the selection rules governing the excitation of surface and bulk plasmon modes and relate the increased blueshift at larger impact parameters to the dominant excitation of higher-order CBP modes.
Importantly, our analytical EELS expression explicitly separates the surface, bulk, and Begrenzung contributions \cite{Ritchie_1957_PhysRev}, thereby providing a transparent physical interpretation of each term within the nonlocal framework.
Altogether, the present results establish a unified analytical foundation for understanding CBP excitation mechanisms and enable comprehensive and accurate interpretation of EEL spectra of small metallic nanoparticles in the bulk plasmon energy range.

\section{Theoretical framework}\label{sec:2}

\begin{figure}[t!]
 \centering
    \includegraphics[width=0.35\textwidth]{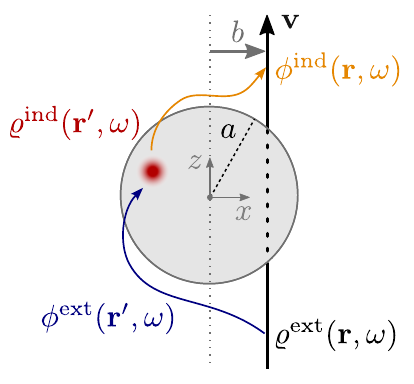}
    \caption{Sketch of a metallic spherical nanoparticle of radius $a$ targeted by a penetrating electron beam with constant impact parameter $b$ and velocity $\textbf{v}$. The electron probe, described by the charge density, $\varrho^{\textrm{ext}}(\textbf{r}, \omega)$, creates a potential, $\phi^{\textrm{ext}}(\textbf{r}',\omega)$.
    This induces a charge density in the nanoparticle, $\varrho^{\textrm{ind}}(\textbf{r}',\omega)$, which generates an induced potential, $\phi^{\textrm{ind}}(\textbf{r},\omega)$, that acts back on the electron probe making it to lose energy.
    }
\label{fig:1}
\end{figure}

The electron beam is modeled as a pointlike charged particle traveling with constant velocity $v$ (nonrecoil approximation) along a trajectory parallel to the $z$ axis at constant distance $b$ (impact parameter) from the center of the nanoparticle of radius $a$, as shown in Fig.~\ref{fig:1}.
The beam traverses the nanoparticle when $b<a$ (penetrating trajectory) and is aloof for $b>a$ (external trajectory).
The charge density associated with the traveling electron can be expressed as \hbox{$\varrho^{\textrm{ext}}(\textbf{r},t) = - e \delta(x-b) \delta(y) \delta(z-vt)$}, where $e$ is the electron charge.
In spherical coordinates, $(r,\theta,\varphi)$, and in the frequency domain, $\omega$, the corresponding expression reads as
\begin{align}
\varrho^{\textrm{ext}}(\textbf{r}, \omega) = - \frac{e}{v} \frac{\delta(r \sin \theta - b) \delta(\varphi)}{r \sin \theta} e^{i \omega r \cos \theta / v}. \label{eq:extchargedens}
\end{align}
The electron probe, described by the external electron density $\varrho^{\textrm{ext}}(\textbf{r}, \omega)$, interacts with the nanoparticle electron gas through its electric potential $\phi^{\textrm{ext}}(\textbf{r}', \omega)$ and induces a charge density $\varrho^{\textrm{ind}}(\textbf{r}',\omega)$ in the nanoparticle.
The induced charge density produces an induced potential $\phi^{\textrm{ind}}(\textbf{r},\omega)$, which acts back on the electron probe, decelerating it and causing it to lose energy (see Fig.~\ref{fig:1}).
Within the quasistatic approximation, the total energy loss experienced by the electron probe is related to the induced potential at the position of the traveling electron, \hbox{$\textbf{r} = (x,y,z) = (b,0,vt)$}, as
\begin{align}
W = \frac{e}{2 \pi} \int^{\infty}_{-\infty} d \omega \int ^{\infty}_{- \infty} dz \frac{\partial \phi^{\textrm{ind}}(\textbf{r}',\omega)}{\partial z'} \bigg{\vert}_{\textbf{r}' = \textbf{r}} e^{- i \omega z / v}. \label{eq:lossgeneral}
\end{align}
The total energy loss is usually expressed as a function of the probability of losing energy $\hbar \omega$ as
\begin{align}
  W = \int^{\infty}_{0} d \omega \, \hbar \, \omega \, \Gamma_{\textrm{EELS}}(\omega),\label{eq:lossprobgeneral}
\end{align}
where $\Gamma_{\textrm{EELS}}(\omega)$ is the electron energy loss probability per energy unit, which is the magnitude directly comparable with the experimentally measured energy loss spectrum.
For finite targets, such as isolated nanoparticles, the EEL probability can be written as \bibnote{By integrating Eq.~\eqref{eq:lossgeneral} by parts, taking into account that the induced potential tends to zero at an infinite distance, $\phi(x,y, \pm \infty) \rightarrow 0$, and that both the induced charge density and induced potential are real variables in the time domain, $\varrho(\textbf{r},\omega) = \varrho^{*}(\textbf{r},-\omega)$ and $\phi(\textbf{r},\omega) = \phi^{*}(\textbf{r},-\omega)$, we obtain Eq.~\eqref{eq:eelsprobgeneral}.}
\begin{align}
    \Gamma_{\textrm{EELS}}(\omega) = \frac{e}{\pi \hbar v} \int_{-\infty}^{\infty} dz \, \textrm{Im} \bigg{\{} - \phi^{\textrm{ind}}(\textbf{r},\omega) e^{-i \omega z/v} \bigg{\}}. \label{eq:eelsprobgeneral}
\end{align}
Thus, to obtain $\Gamma_{\textrm{EELS}}(\omega)$, we need to compute the induced scalar potential $\phi^{\textrm{ind}}(\textbf{r}, \omega)$ along the trajectory of the electron beam, which may be calculated assuming different approaches to address the probe-target interaction \cite{Rivacoba_1992_PhysRevLett, Zabala_1993_PhysRevB, Wang_1996_Micron, Rivacoba_2000_ProgrSurfSci, GarciadeAbajo_2010_RevModPhys}.
In the next subsections, we derive the induced scalar potential from the charge density $\varrho^{\textrm{ind}}(\textbf{r}, \omega)$ induced at the spherical metallic nanoparticle as a response to the electron beam using a linear hydrodynamic model.

\subsection{Dynamics of an electron gas confined to a spherical nanoparticle} \label{sec:theory_HM}

To describe the response of a metallic nanoparticle to a penetrating fast electron, we model the conduction electrons as a homogeneous free‑electron gas of equilibrium density \hbox{$n = 3/(4\pi r_s^3)$} confined within a sphere of radius $a$.
The collective dynamics of this electron gas is treated within the linear hydrodynamic model, which incorporates spatial dispersion through the finite compressibility of the electron gas.
Throughout this work, we employ atomic units ($\hbar = e = m_{e} = 4 \pi \varepsilon_{0} = 1$).
The linearized continuity and force‑balance equations for the electron gas lead to a closed equation for the induced charge density \cite{Rivacoba_2019_Ultramicroscopy}, which inside the metal satisfies the inhomogeneous Helmholtz equation:
\begin{align}
[\nabla^{2} + \mu^{2}(\omega)] \varrho^{\textrm{ind}}(\textbf{r}, \omega) = \frac{\omega_{p}^{2}}{\beta^{2}} \varrho^{\textrm{ext}}(\textbf{r}, \omega) , \label{eq:Helmholtz}
\end{align}
where the longitudinal hydrodynamic wave vector $\mu(\omega)$ satisfies
\begin{align}
\mu^{2}(\omega) = \frac{1}{\beta^{2}} \Big{[} \omega (\omega + i \gamma) - \omega_{p}^{2} \Big{]}, \label{eq:mufunction}
\end{align}
which governs the oscillation and decay length of longitudinal modes inside the nanoparticle.
Here, $\omega_p=\sqrt{4\pi n}$ is the plasma frequency, $\gamma$ is a phenomenological damping rate, and $\beta = \sqrt{3/5} v_{\textrm{F}}$ is the hydrodynamic dispersion parameter determined by the Fermi velocity of the electron gas $v_{\textrm{F}} = (3\pi^2n)^{1/3}$.
Notice that $1/\beta^{2}$ is proportional to the compressibility of the electron gas ($\beta \rightarrow 0$ in the LRA) \cite{Raza_2015_NatComm, Boardman_1982_book}.
The hydrodynamic boundary condition of a vanishing normal component of the electron current at the particle surface yields \cite{Crowell_1968_PhysRev, VilloPerez_2009_SurfSci}
\begin{align}
-\beta^2\frac{\partial \varrho^{\textrm{ind}}(\textbf{r},\omega)}{\partial r} =  \frac{1}{4\pi} \omega_p^2 \frac{\partial \phi(\textbf{r}, \omega)}{\partial r}.
\label{eq:BC_hydro}
\end{align}
This condition relates the normal derivative of the induced charge density to that of the total scalar potential $\phi(\textbf{r}, \omega)$ and introduces the physical constraint that electrons do not escape the nanoparticle, leading to the confinement of longitudinal oscillations.

When a background dielectric response is present --either inside the nanoparticle through a core‑polarization term $\varepsilon_\infty$ or outside through a host medium of permittivity $\varepsilon_m$-- the hydrodynamic problem must be solved consistently, together with Poisson’s equation for the scalar potential inside and outside the nanoparticle, and with the customary electromagnetic boundary conditions at the interface.
These conditions ensure the proper matching of the fields and correctly incorporate the influence of the background permittivities \cite{V.Baltz_1995_ZPhysB}.

Instead, in the present work we restrict our analysis to the case ($\varepsilon_\infty = \varepsilon_m = 1$), for which the potential is fully determined by the Coulomb potential
\begin{align}
\phi(\textbf{r}, \omega) & = \int \frac{\varrho(\textbf{r}', \omega) }{|\textbf{r} - \textbf{r}'|} d\textbf{r}', \label{eq:potential_integral}
\end{align}
with $\varrho(\textbf{r}, \omega) = \varrho^{\textrm{ind}}(\textbf{r}, \omega) + \varrho^{\textrm{ext}}(\textbf{r}, \omega)$.
This assumption reduces the problem to solving Eq.~\eqref{eq:Helmholtz} with the hydrodynamic boundary condition given by Eq.~\eqref{eq:BC_hydro}, exclusively written in terms of the external and induced charge densities:
\begin{align}
\nonumber \hspace{2em}&\hspace{-2em} - \frac{1}{4\pi}\omega_p^2 \frac{\partial}{\partial r}\int d\textbf{r}'\frac{\varrho^{\textrm{ind}}(\textbf{r}',\omega)}{\vert \textbf{r}'-\textbf{r}\vert}-\beta^2\frac{\partial\varrho^{\textrm{ind}}(\textbf{r},\omega) }{\partial r}= \\
= & \frac{1}{4\pi} \omega_p^2 \frac{\partial}{\partial r}\int d\textbf{r}'\frac{\varrho^{\textrm{ext}}(\textbf{r}',\omega)}{\vert \textbf{r}'-\textbf{r}\vert}. \label{eq:BC}
\end{align}
The solution of Eq.~\eqref{eq:Helmholtz} with Eq.~\eqref{eq:BC} determines the induced potential through Eq.~\eqref{eq:potential_integral}, and sets the framework for the multipolar expansion and analytical expression of the electron energy loss probability derived in the next subsection.

\subsection{Electron energy loss probability for electron trajectories penetrating a spherical nanoparticle}\label{sec:eloss}

We explain in Appendix \ref{app:a} the steps followed to obtain the solution of the inhomogeneous Helmholtz equation \eqref{eq:Helmholtz} inside the nanoparticle, which we outline here.
For the sake of brevity, we omit the explicit $\omega$ dependence in the following expressions.
Making use of the spherical symmetry of the target, we expand the external, $\varrho^{\textrm{ext}}(\textbf{r})$ [see Eq.~\eqref{eq:extchargedens}], and the induced, $\varrho^{\textrm{ind}}(\textbf{r})$, charge densities in terms of the spherical harmonics, $Y_{l}^{m}(\Omega)$, and introduce the radial components of the induced and external charge densities, $\varrho^{\textrm{ind}}_{lm}(r)$ and $\varrho_{lm}^{\textrm{ext}}(r)$, respectively [see Eqs.~\eqref{eq:chargedensityextgeneral} and \eqref{eq:chargedensitygeneral}].
Notice that the radial component of the solution, $\varrho^{\textrm{ind}}_{lm}(r)$, satisfies the spherical Bessel differential equation [Eq.~\eqref{eq:bessel_differential}].
By expanding the Coulomb potentials of the boundary condition [Eq.~\eqref{eq:BC}] into spherical harmonics, we obtain Eq.~\eqref{eq:BCspherical}, the boundary condition in terms of $\varrho^{\textrm{ind}}_{lm}(r)$ and $\varrho^{\textrm{ext}}_{lm}(r)$.

After some algebra, the radial component of the induced charge density that satisfies both the boundary condition in Eq.~\eqref{eq:BCspherical} and the inhomogeneous Helmholtz equation [Eq.~\eqref{eq:Helmholtz}] results in
\begin{align}
\nonumber \varrho^{\textrm{ind}}_{lm}(r) = & \, \frac{2 \omega_p^2 \alpha_{lm}}{v}  \bigg{\{} \bigg{[} \frac{(l+1) \mu}{(2l+1)a M_{l}} \bigg{(} 1 + \frac{\omega^2_p}{\mu^2 \beta^2} \bigg{)}\mathcal{I}_{lm}(b,a) \\
\nonumber & - \frac{l\,\mu}{(2l+1) a M_{l}} \mathcal{O}_{lm}(a,\infty) + \frac{N_{l}}{M_{l}} \frac{1}{\beta^2} \mathcal{J}_{lm}(b,a) \\
\nonumber & - \frac{1}{\beta^2} \mathcal{Y}_{lm}(b,a) \bigg{]} j_{l}(\mu r) - \frac{1}{\beta^2} \Theta \Big{(} \frac{r}{b} - 1 \Big{)} \\
& \times\bigg{[} y_{l}(\mu r) \mathcal{J}_{lm}(b,r) - j_{l}(\mu r) \mathcal{Y}_{lm}(b,r) \bigg{]} \bigg{\}}, \label{eq:chargedensitytotal}
\end{align}
where,
\begin{subequations}
\begin{align}
N_{l} = \omega_p^2 \frac{l+1}{2l+1} y_{l+1}(\mu a) - \beta^2 \mu^2 y'_{l}(\mu a), \label{eq:Nl}\\
M_{l} = \omega_p^2 \frac{l+1}{2l+1} j_{l+1}(\mu a) - \beta^2 \mu^2 j'_{l}(\mu a). \label{eq:Ml}
\end{align}
\end{subequations}
Here, $j_{l}(x)$ and $y_{l}(x)$ are the spherical Bessel functions of the first and second kind, respectively, $\Theta (x)$ is the Heaviside step function, and the following coefficients and functions are introduced in Appendix \ref{app:a}: $\alpha_{lm}$, $\mathcal{I}_{lm}$, $\mathcal{O}_{lm}$, $\mathcal{J}_{lm}$, and $\mathcal{Y}_{lm}$ [see Eqs.~\eqref{eq:alphalm} and \eqref{eq:Ilm}-\eqref{eq:Ylm}].

Once the induced charge density is known, the induced potential can be obtained via the Coulomb potential in Eq.~\eqref{eq:potential_integral}, which is expanded in spherical harmonics,
\begin{align}
\phi^{\textrm{ind}}(\textbf{r}) = \sum_{l=0}^{\infty} \sum_{m=-l}^{l} Y_{l}^{m}(\Omega) \phi^{\textrm{ind}}_{lm} (r), \label{eq:indpotsphharm}
\end{align}
with the radial component of the induced potential inside and outside the nanoparticle given by:
\begin{align}
\phi^{\textrm{ind}}_{lm}(r) & = \frac{4 \pi}{2 l + 1} \int_{0}^{a} dr' \, r'^2 \frac{r_{<}^{l}}{r_{>}^{l+1}} \varrho^{\textrm{ind}}_{lm}(r'),\label{eq:indporadial}
\end{align}
where $r_{<}=\min(r,r')$ and $r_{>}=\max(r,r')$.
After some involved but elementary calculations (see Appendix \ref{app:b}), one obtains for $r>a$,
\begin{align}
\nonumber \phi_{lm}^{\textrm{ind},o}(r) & =  8 \pi \frac{\omega_p^2}{v} \frac{l \, \alpha_{lm}}{(2l +1) \,M_{l}}\bigg{\{} \frac{j_{l-1}(\mu a)}{2l+1} \mathcal{I}_{lm}(b,a) \\
\nonumber & \hspace{6em} - \frac{j_{l+1}(\mu a)}{2l+1} \, \mathcal{O}_{lm}(a,\infty) \\
& \hspace{6em} - \frac{1}{\mu^2 a^2} \mathcal{J}_{lm}(b,a) \bigg{\}} \bigg{(} \frac{a}{r} \bigg{)}^{l+1}, \label{eq:indpotout}
\end{align}
\begin{widetext}
and for $r<a$,
\begin{align}
\nonumber \phi^{\textrm{ind},i}_{lm}(r) = & \; 8 \pi \frac{\omega_p^2}{v} \frac{\alpha_{lm}}{2l +1} \bigg{\{} \bigg{[} \frac{l+1}{\mu a M_{l}} \bigg{(} 1 + \frac{\omega^2_p}{\mu^2 \beta^2} \bigg{)} \mathcal{I}_{lm}(b,a) - \frac{l}{\mu a M_{l}} \mathcal{O}_{lm}(a,\infty) \\
\nonumber & + \frac{2l+1}{\mu^{2}\beta^2 M_{l}} [N_{l} \mathcal{J}_{lm}(b,a) - M_{l} \mathcal{Y}_{lm}(b,a)] \bigg{]} j_{l}(\mu r)  \\
\nonumber & + \bigg{[} - \frac{(l+1) j_{l-1}(\mu a)}{(2l+1)M_{l}} \bigg{(} 1 + \frac{\omega^2_p}{\mu^2 \beta^2} \bigg{)} \mathcal{I}_{lm}(b,a) + \frac{l \, j_{l-1}(\mu a)}{(2l+1) M_{l}} \mathcal{O}_{lm}(a,\infty) - \frac{1}{\mu^2 \beta^2} \mathcal{O}_{lm}(b,a) \\
\nonumber & + \frac{l+1}{\mu^2 a^2 M_{l}} \bigg{(} 1 + \frac{\omega^2_p}{\mu^2 \beta^2} \bigg{)} \mathcal{J}_{lm}(b,a) \bigg{]} \bigg{(} \frac{r}{a} \bigg{)} ^{l}\\
\nonumber & - \Theta \Big{(} \frac{r}{b} - 1 \Big{)} \frac{2l+1}{\mu^2 \beta^2} \bigg{[} y_{l}(\mu r) \mathcal{J}_{lm}(b,r) - j_{l}(\mu r) \mathcal{Y}_{lm}(b,r) \\
& + \frac{\mathcal{I}_{lm}(b,r)}{2l+1} \bigg{(} \frac{a}{r} \bigg{)}^{l+1} - \frac{\mathcal{O}_{lm}(b,r)}{2l+1} \bigg{(} \frac{r}{a} \bigg{)}^{l} \bigg{]} \bigg{\}}. \label{eq:indpotin}
\end{align}
This expansion of the induced potential in spherical harmonics allows us to write the EEL probability [Eq.~\eqref{eq:eelsprobgeneral}] as the contributions of $(l,m)$ multipoles:
\begin{align}
\Gamma_{\textrm{EELS}} = & \; \frac{2}{\pi v} \sum_{l=0}^{\infty} \sum_{m=-l}^{l} \alpha_{lm} (-1)^{l+m+1} \int^{\infty}_{0} dz \, \textrm{Im} \bigg{[} \phi^{\textrm{ind}}_{lm}(r) P_{l}^{m} \bigg{(} \frac{z}{r} \bigg{)} g_{lm}\bigg{(}\frac{\omega z}{v} \bigg{)} \bigg{]}, \label{eq:eelsspherical}
\end{align}
where $r = \sqrt{z^2 + b^2}$, $P_{l}^{m}(x)$ is the associated Legendre polynomial of degree $l$ and order $m$, and $g_{lm}(x)$ is defined by Eq.~\eqref{eq:glm}.
We also take into account that $P_{l}^{m}(-x) = (-1)^{l+m} P_{l}^{m}(x)$ and $g^{*}_{lm}(x) = (-1)^{l+m} g_{lm}(x)$.
Introducing the potentials of Eqs.~\eqref{eq:indpotout} and \eqref{eq:indpotin} into Eq.~\eqref{eq:eelsspherical}, and after some algebra (see Appendix \ref{app:c}), one obtains finally the EEL probability as the contribution of three terms,
\begin{align}
\Gamma_{\textrm{EELS}} = \Gamma^{\textrm{bulk}} + \Gamma^{\textrm{Begr}} + \Gamma^{\textrm{ext}}, \label{eq:EELprobabilitygen}
\end{align}
where
\begin{subequations}
\begin{align}
\Gamma^{\textrm{bulk}} = & \, \frac{8 e^{2} \omega_p^2}{\pi \hbar v^2} \sum_{l=0}^{\infty} \sum_{m=0}^{l} \chi_{lm} \; \textrm{Im} \Bigg{\{} \frac{1}{\mu^2 \beta^2} \mathcal{F}_{lm} \Bigg{\}}, \label{eq:Gammabulk} \\
\Gamma^{\textrm{Begr}} = & \, \Gamma^{\textrm{Begr,i}} + \Gamma^{\textrm{Begr,o}}, \label{eq:GammaBegr} \\
\nonumber \Gamma^{\textrm{Begr,i}} = & \, \frac{4 e^{2} a \omega_p^2}{\pi \hbar v^2} \sum_{l=0}^{\infty} \sum_{m=0}^{l} \chi_{lm} \; \textrm{Im} \Bigg{\{} \frac{l+1}{M_{l}} \bigg{(} 1 + \frac{\omega^2_p}{\mu^2 \beta^2} \bigg{)} \mathcal{I}_{lm}(b,a) \bigg{[} \frac{j_{l-1}(\mu a)}{(2l+1)} \mathcal{I}_{lm}(b,a) - \frac{2}{\mu^{2} a^{2}} \mathcal{J}_{lm}(b,a) \bigg{]} \\
& \, - \frac{2l+1}{\mu^{3} \beta^{2} a} \bigg{[} \frac{N_{l}}{M_{l}} [\mathcal{J}_{lm}(b,a)]^2 - 2 \, \mathcal{H}_{lm} \bigg{]} \Bigg{\}}, \label{eq:GammaBegr_i} \\
\Gamma^{\textrm{Begr,o}} = & \frac{4 e^{2} a \omega_p^2}{\pi \hbar v^2} \sum_{l=0}^{\infty} \sum_{m=0}^{l} \chi_{lm} \; \textrm{Im} \Bigg{\{} \frac{2l}{M_{l}} \mathcal{O}_{lm}(a, \infty) \bigg{[} - \frac{j_{l-1}(\mu a)}{(2l+1)} \mathcal{I}_{lm}(b,a) + 
\frac{1}{\mu^{2} a^{2}} \mathcal{J}_{lm}(b,a) \bigg{]} \Bigg{\}}, \label{eq:GammaBegr_o} \\
\Gamma^{\textrm{ext}} = & \, \frac{4 e^{2} a \omega_p^2}{\pi \hbar v^2} \sum_{l=0}^{\infty} \sum_{m=0}^{l} \chi_{lm} \; \textrm{Im} \Bigg{\{} \frac{l \, j_{l+1}(\mu a)}{(2l+1) M_{l}} [\mathcal{O}_{lm}(a,\infty)]^2 \Bigg{\}}, \label{eq:Gammaext}
\end{align}
\end{subequations}
\end{widetext}
with $\mathcal{F}_{lm}$ and $\mathcal{H}_{lm}$ defined by Eqs.~\eqref{eq:Flm} and \eqref{eq:Hlm}, and the prefactor,
\begin{align}
    \chi_{lm} = (-1)^{l+m} (2 - \delta_{m0}) \frac{(l-m)!}{(l+m)!}. \label{eq:prefactor_eloss}
\end{align}
We explicitly include the electron charge $e$ and the reduced Planck's constant $\hbar$ in these expressions.

\begin{figure*}[t!]
 \centering
    \includegraphics[width=1\textwidth]{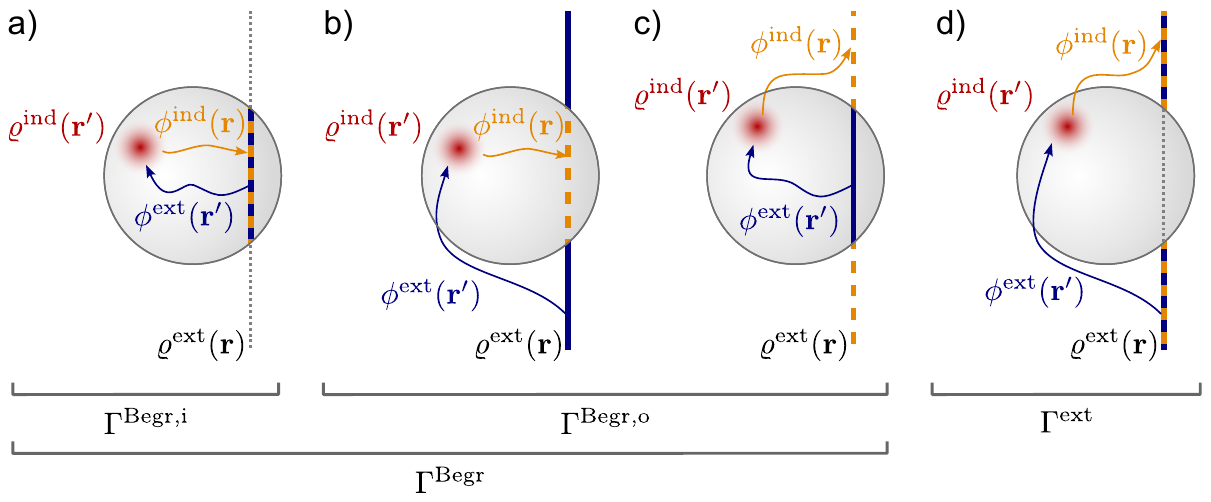}
    \caption{Schematic illustration of the contributions to the total electron energy loss probability $\Gamma_{\textrm{EELS}}$.
    The solid blue line represents the external charge density, $\varrho^{\textrm{ext}}(\textbf{r})$, which generates an external potential (blue arrow), $\phi^{\textrm{ext}}(\textbf{r}')$.
    This potential induces a charge density (red spot), $\varrho^{\textrm{ind}}(\textbf{r}')$, within the nanoparticle.
    In turn, the induced charge density creates an induced potential (orange arrow), $\phi^{\textrm{ind}}(\textbf{r})$, that acts back on the external charge density (dashed orange line) resulting in the energy loss.
    The figure outlines the different contributions to $\Gamma_{\textrm{EELS}}$, based on the location of the electron beam path -inside or outside the nanoparticle- during both the excitation (solid blue line) and the energy loss (dashed orange line) processes:
    (a) inside-inside, both excitation and energy loss occur within the nanoparticle (contributes to $\Gamma^{\textrm{Begr,i}}$);
    (b) outside–inside, excitation occurs outside, while energy loss occurs inside the nanoparticle;
    (c) inside–outside, excitation occurs inside, while energy loss occurs outside the nanoparticle
    [both panels (b) and (c) contribute to $\Gamma^{\textrm{Begr,o}}$];
    and d) outside–outside, both excitation and energy loss occur outside the nanoparticle (contributes to $\Gamma^{\textrm{ext}}$).
    }
\label{fig:2}
\end{figure*}

Here, the first term, $\Gamma^{\textrm{bulk}}$, related to the bulk losses in an unbound medium, is reduced by the Begrenzung term $\Gamma^{\textrm{Begr}}$ that accounts for the presence of the boundary \cite{Ritchie_1957_PhysRev}.
This Begrenzung term can be split into two terms depending on the position of the electron probe: the \textit{inner} term, $\Gamma^{\textrm{Begr,i}}$, and the \textit{outer} term, $\Gamma^{\textrm{Begr,o}}$.
We sketch in Fig. \ref{fig:2} the physical interpretation of each term.
The solid blue line depicts the segment of the electron beam path corresponding to 
the external charge density, $\varrho^\textrm{ext}(\textbf{r})$, which produces the external potential, $\phi^{\textrm{ext}}(\textbf{r}')$, that induces $\varrho^{\textrm{ind}}(\textbf{r}')$ at the nanoparticle.
The dashed orange line depicts the segment of the electron beam path on which the induced potential, $\phi^{\textrm{ind}}(\textbf{r}')$, produced by $\varrho^{\textrm{ind}}(\textbf{r}')$, acts back producing the subsequent energy loss.
Therefore, the \textit{inner} term, $\Gamma^{\textrm{Begr,i}}$, accounts for the losses when the electron probe is inside the nanoparticle, both when perturbing the electron cloud and undergoing the energy loss, as illustrated in Fig.~\hyperref[fig:2]{\ref*{fig:2}(a)}.
On the other hand, the \textit{outer} term, $\Gamma^{\textrm{Begr,o}}$, is associated with two distinct configurations: in Fig.~\hyperref[fig:2]{\ref*{fig:2}(b)} the electron probe is outside the nanoparticle when exciting it, and inside when undergoing the energy loss, and in Fig.~\hyperref[fig:2]{\ref*{fig:2}(c)} the electron probe is inside the nanoparticle when exciting it, and outside when undergoing the energy loss.
The last term $\Gamma^{\textrm{ext}}$ contains the contribution associated with the electron probe outside the nanoparticle, both when exciting it and when experiencing the energy loss, as sketched in Fig.~\hyperref[fig:2]{\ref*{fig:2}(d)}.
The main contribution in this term is associated with LSPs, although there is also a small contribution coming from certain CBPs (only observable if a logarithmic scale is used \cite{Christensen_2014_ACSNano}).
Notably, the contributions of the terms in the second row of $\Gamma^{\textrm{Begr,i}}$ in Eq.~\eqref{eq:GammaBegr_i} are associated with the excitation of CBPs, while the rest of the terms have mixed contributions from both LSPs and CBPs.

For external trajectories ($b>a$), only the term $\Gamma^{\textrm{ext}}$ contributes to $\Gamma_{\textrm{EELS}}$ in Eq.~\eqref{eq:EELprobabilitygen} [with $\mathcal{O}_{lm}(b, \infty)$ instead of $\mathcal{O}_{lm}(a, \infty)$], which reduces to \cite{TranThoai_1986_physstatussolidib}
\begin{align}
    \nonumber \Gamma_{\textrm{EELS}} = \, & \frac{4 e^{2} a \omega_p^2}{\pi \hbar v^2} \sum_{l=0}^{\infty} \sum_{m=0}^{l} \frac{(2 - \delta_{m0})}{(l-m)!(l+m)!} \bigg{(} \frac{\omega a}{v} \bigg{)}^{2l}\\
    & \times K_{m}^{2} \bigg{(} \frac{\omega b}{v} \bigg{)} \textrm{Im} \bigg{\{} \frac{l \, j_{l+1}(\mu a)}{(2l+1) M_{l}} \bigg{\}}, \label{eq:EELprobabilityext}
\end{align}
where $K_{m}$ is the modified Bessel function of order $m$.

\subsection{Eigenmodes of the spherical nanoparticle}

\begin{figure}[t!]
 \centering
    \includegraphics[width=0.45\textwidth]{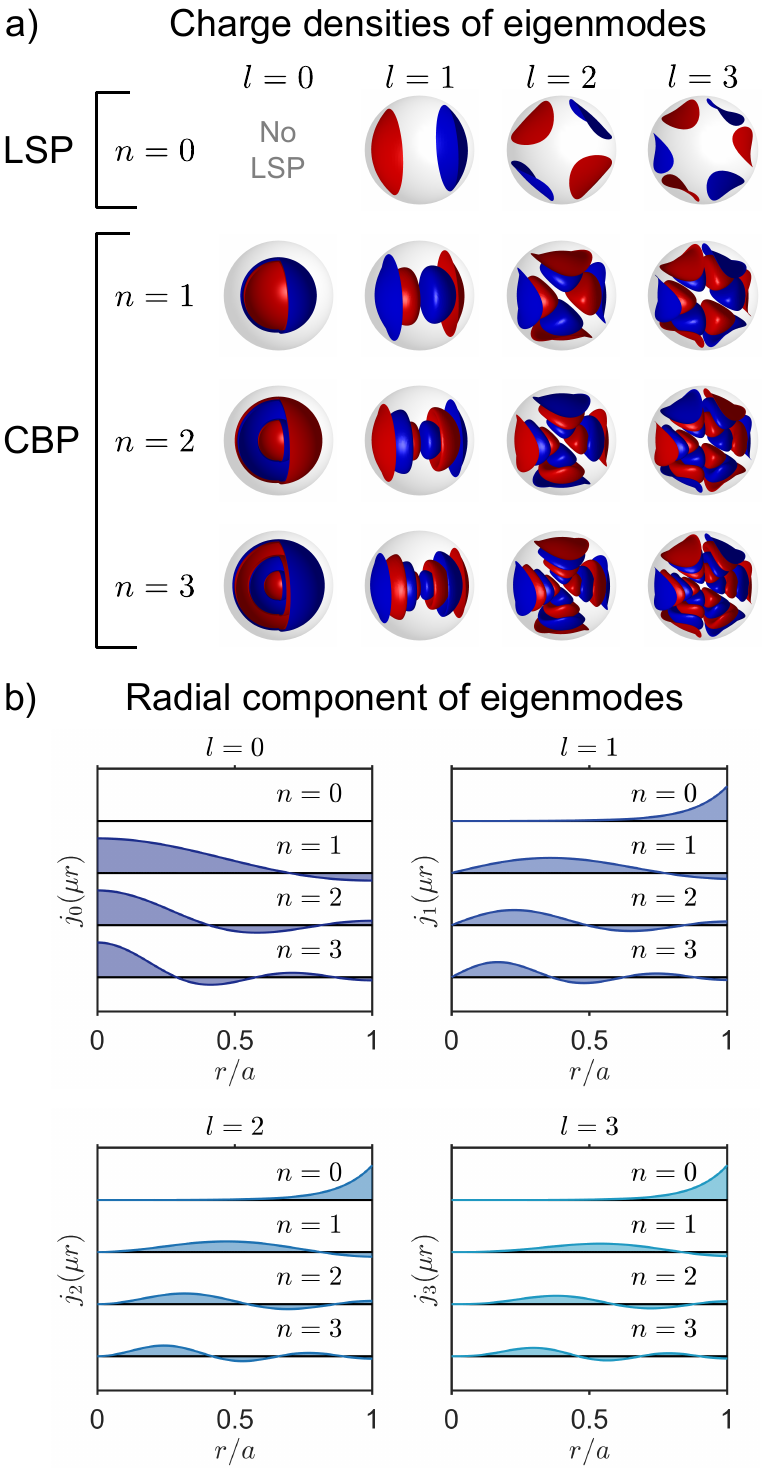}
    \caption{(a) Charge density isosurfaces associated with the lowest order eigenmodes of a spherical nanoparticle obtained from Eq.~\eqref{eq:modecond}. The number $l$ is associated with the polar symmetry and $n$ to the number of nodes of the charge density in the radial direction corresponding to the mode. Red and blue surfaces show charge density of opposing phase. The contour of the spherical nanoparticle is highlighted in gray. (b) Radial component [$j_{l}(\mu r)$, spherical Bessel function of the first kind] associated with the charge density of the eigenmodes, as a function of the reduced radius $r/a$.
    }
\label{fig:3}
\end{figure}

Within the hydrodynamic model described in the previous subsection, the eigenmodes of a free-electron gas confined to a spherical nanoparticle are obtained from the solution of the homogeneous problem [Eqs.~\eqref{eq:Helmholtz} and \eqref{eq:BC}], with $\varrho^{\textrm{ext}} (\textbf{r}, \omega) = 0$.
This leads to the mode condition $M_{l}=0$ [see Eq.~\eqref{eq:Ml}], well known in the literature \cite{Fujimoto_1968_JPhysSocJpn, TranThoai_1986_physstatussolidib, V.Baltz_1995_ZPhysB}:
\begin{align}
\omega_p^2\frac{l+1}{2l+1} j_{l+1}(\mu a) - \beta^2 \mu^2 j'_{l}(\mu a) = 0. \label{eq:modecond}
\end{align}
Notice that this expression includes nonlocal effects but does not account for relativistic effects (quasistatic limit).

The dispersion from Eq.~\eqref{eq:modecond} does not break the degeneracy with respect to the azimuthal number $m$, in analogy to the LRA.
Nevertheless, it gives multiple eigenenergies $\omega_{ln}$, labeled as \hbox{$n=0,1,2,3,...$}, and associated with the number of nodes of the charge density in the radial direction, for each $l$ polar term.
These solutions are obtained numerically and provide all the possible plasma resonances excited at a spherical nanoparticle of radius $a$ for a homogeneous electron gas with Wigner-Seitz radius $r_{s}$.
In the case of $l=0$, their energies are obtained analytically as:
\begin{align}
\frac{\omega_{0n}^{2}}{\omega_{p}^{2}} = 1 + \frac{x_{0,n}^{2}}{\omega_{p}^{2}} \bigg{(} \frac{\beta}{a} \bigg{)}^{2} , \label{eq:modecondzero}
\end{align}
where $x_{0,n}$ is the $n$th positive root of $j'_{0}(x)$, the derivative of the zeroth-order spherical Bessel function (assuming \hbox{$\gamma=0$)}.

It is illustrative to make a distinction between the first solution for each $l$ ($n=0$) and the rest of the solutions ($n>0$).
The former represents the LSPs and the latter the CBPs (see Ref.~\cite{Christensen_2014_ACSNano} for details in this regard), except for the nonphysical $(l,n) = (0,0)$ mode, which violates the charge conservation principle.
This distinction is clearly visualized in Fig.~\hyperref[fig:3]{\ref*{fig:3}(a)}, where we show the charge density distributions associated with the lowest order eigenmodes ($l=0\hbox{ - }3$, $m=l$, $n=0\hbox{ - }3$). 
The radial components of the charge density associated with the modes inside the nanoparticle, $j_{l}(\mu r)$, for $l=0 \hbox{ - }3$ and $n =0 \hbox{ - }3$ (LSPs top row and CBPs bottom three rows) are represented in Fig.~\hyperref[fig:3]{\ref*{fig:3}(b)}.
These modes clearly show a distinct nature: LSPs ($n=0$) have an exponential decay from the surface into the nanoparticle, while the CBPs ($n>0$) present $n$ oscillations of the charge density inside the nanoparticle.
As we describe in the next section, the excitation of these modes by the penetrating electron probe can be understood in terms of their symmetry, characterized by the $(l,m,n)$ tuples.
Owing to the energy degeneracy, we omit $m$ hereafter and simply refer to the modes as $(l,n)$, unless stated otherwise.

\subsection{Comparison with the local response approximation} \label{sec:LRA}
Within the LRA limit, neglecting nonlocal effects, there is no dispersion of the modes, i.e., $\beta \rightarrow 0$ and $\mu a \rightarrow \infty$.
Then, making use of the large argument limits of the spherical Bessel functions, the mode condition given by Eq.~\eqref{eq:modecond} reduces to
\begin{align}
    \omega = \sqrt{\frac{l+1}{2l+1}} \omega_{p}
\end{align}
which gives the well-known LSP energies in spherical nanoparticles in the quasistatic limit. 
Notice that the limit $\mu a \rightarrow \infty$ washes out the CBPs from the response of the system.
In this limit, Eq.~\eqref{eq:EELprobabilityext} recovers the expression given by Ferrell and Echenique for external trajectories \cite{Ferrell_1985_PhysRevLett}.
For penetrating trajectories, and defining \hbox{$\varepsilon = 1 - \omega_{p}^{2}/[\omega(\omega + i \gamma)]$},  Eqs.~\eqref{eq:Gammabulk}-\eqref{eq:Gammaext} reduce to the following expressions:
\begin{subequations}
\begin{align}
    \Gamma^{\textrm{bulk}} = & \, \frac{8 e^{2}}{\pi \hbar v^2} \sum_{l=0}^{\infty} \sum_{m=0}^{l} \chi_{lm} \; \textrm{Im} \bigg{\{} - \frac{1}{\varepsilon} \mathcal{F}_{lm}\bigg{\}}, \label{eq:Gammabulk_local} \\
    \Gamma^{\textrm{Begr,i}} & = \frac{4 e^{2} a}{\pi \hbar v^2} \sum_{l=0}^{\infty} \sum_{m=0}^{l} \chi_{lm} \; \textrm{Im} \Big{\{} \eta_{l} \; [\mathcal{I}_{lm}(b,a)]^{2} \Big{\}}, \label{eq:GammaBegr_i_local}\\
    \nonumber \Gamma^{\textrm{Begr,o}} & = \frac{4 e^{2} a}{\pi \hbar v^2} \sum_{l=0}^{\infty} \sum_{m=0}^{l} \chi_{lm} \; \\
    \hspace{-4em} & \hspace{4em} \times \textrm{Im} \Big{\{} 2 \zeta_{l} \; \mathcal{I}_{lm}(b,a) \mathcal{O}_{lm}(a,\infty) \Big{\}}, \label{eq:GammaBegr_o_local}\\
    \nonumber \Gamma^{\textrm{ext}} & = \frac{4 e^{2} a}{\pi \hbar v^2} \sum_{l=0}^{\infty} \sum_{m=0}^{l} \chi_{lm} \\
    \hspace{-4em} & \hspace{4em} \times \textrm{Im} \Big{\{} \zeta_{l} \; [\mathcal{O}_{lm}(a,\infty)]^2 \Big{\}} \label{eq:Gammaext_local},
\end{align}
\end{subequations}
with $\chi_{lm}$ defined by Eq.~\eqref{eq:prefactor_eloss}, and
\begin{align}
    \eta_{l} & = \frac{(l+1)(\varepsilon-1)}{\varepsilon(l \varepsilon + l +1 )}, \\
    \zeta_{l} & = \frac{l (1 - \varepsilon)}{(l+1 + l\varepsilon)}.
\end{align}
The sum of the Begrenzung and external terms becomes equivalent to the expression for the surface corrections to the EEL probability reported in Refs.~\cite{Echenique1987, TranThoai_1988_physstatussolidia}.
In contrast to the nonlocal Begrenzung terms [Eqs.~\eqref{eq:GammaBegr_i} and \eqref{eq:GammaBegr_o}], their local counterparts [Eqs.~\eqref{eq:GammaBegr_i_local} and \eqref{eq:GammaBegr_o_local}] involve no expressions containing $\mathcal{J}_{lm}$ terms because they vanish in the limit $\mu a \rightarrow \infty$.
This disappearance is directly linked to the physical role of the spherical Bessel functions $j_{l}$ and $y_{l}$, which encapsulate the finite-pressure effects ($\beta \neq 0$) in the hydrodynamic description and underpin the emergence of CBPs.
Setting $\beta = 0$ removes nonlocality and, with it, any trace of the spherical Bessel functions, thereby eliminating CBPs entirely from the LRA expressions.
This distinction becomes evident in the results presented in the next section.

On the other hand, the contribution of the bulk plasmon to the losses diverges when the summation is performed over all $l$ modes.
This happens because, within the LRA, the expression for the EEL probability per unit length in an infinite medium is given by the integral \cite{Ritchie_1957_PhysRev}
\begin{align}
\frac{d \, \Gamma_{\textrm{EELS}}}{dz} = \frac{2e^{2}}{\pi \hbar v^{2}} \int_{0}^{\infty} dQ \frac{Q}{ \Big{(} Q^{2} + \frac{\omega^{2}}{v^{2}} \Big{)}} \textrm{Im} \Bigg{\{} -\frac{1}{\varepsilon} \Bigg{\}}, \label{eq:local_bulk_integral}
\end{align}
where $Q$ denotes the modulus of the wave‑vector component perpendicular to the electron beam trajectory.
This integral diverges for large values of $Q$ because its upper bound extends to $\infty$.
To overcome this divergence, a finite momentum cutoff $q_{c}$ is introduced, typically chosen to match the collection angles used in STEM experiments.
Multiplying Eq.~\eqref{eq:local_bulk_integral} by the distance traveled by the electron inside the metal, $L=\sqrt{a^{2}-b^{2}}$, the loss probability becomes:
\begin{align}
    \Gamma_{\textrm{LRA}}^{\textrm{bulk}} = \frac{2 e^{2}L}{\pi \hbar v^{2}} \textrm{Im} \bigg{\{} - \frac{1}{\varepsilon} \bigg{\}} \ln (q_{c} v/\omega), \label{eq:Gamma_bulk_local_log}
\end{align}
which is the standard expression reported in the literature \cite{GarciadeAbajo_2010_RevModPhys}.
It is worth noting that in Eq. \eqref{eq:Gammabulk_local} the momentum cutoff is effectively implemented through the choice of a maximum value $l_{\textrm{max}}$.
In contrast, within the hydrodynamic model, only a few modes are required to achieve convergence of the total $\Gamma_{\textrm{EELS}}$.

\section{Results} \label{sec:results}
To illustrate the analytical expression obtained within the hydrodynamic model, we analyze the EEL probability of a spherical nanoparticle excited by swift electron beams with penetrating trajectories.
We first consider a sodium nanoparticle ($r_\textrm{s} = 2.08 \textrm{ \AA}$, $\omega_{p} \sim 6.05 \textrm{ eV}$) with radius $a = 1.5 \textrm{ nm}$ and damping term $\gamma = \gamma_{\infty} + \frac{3 \, v_{\textrm{F}}}{4 a}$.
The first term $\gamma_{\infty} = 0.1 \textrm{ eV}$ corresponds to the unbound medium and the second term $3 \, v_{\textrm{F}}/4 a$ accounts for electron surface-scattering effects due to nanoparticle size, which are significant at this small nanoparticle size \cite{Kreibig_1969_ZPhys, Moroz_2008_JPhysChemC}.

\subsection{Excitation of CBPs by penetrating electron beam trajectories}
\begin{figure*}[t!]
 \centering
    \includegraphics[width=1\textwidth]{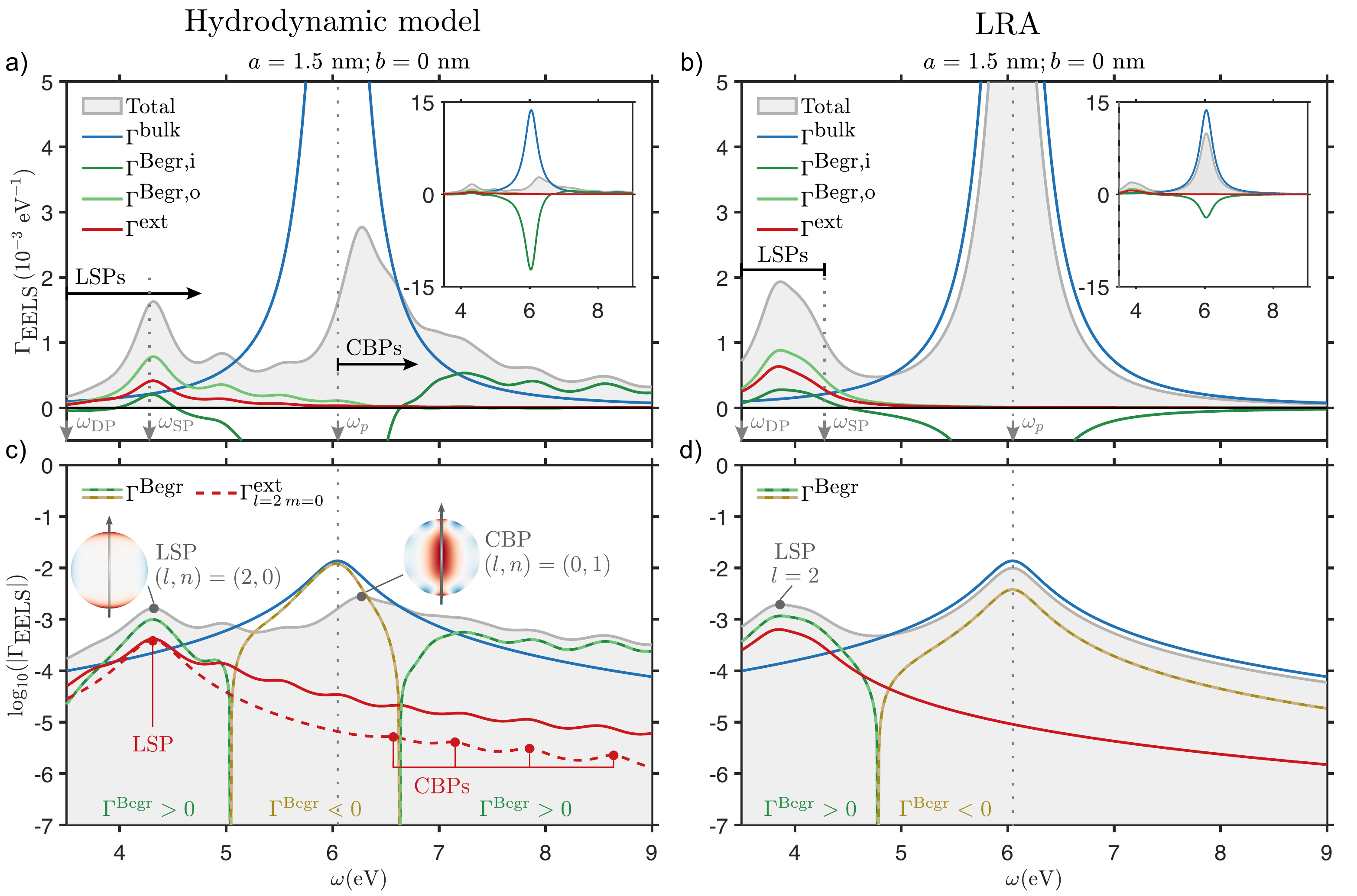}
    \caption{Electron energy loss probability (gray shaded area) and its contributions: the bulk term (blue curve), inner and outer Begrenzung terms (solid dark green and clear green curves, respectively), and external term (red curve), for multipole order cutoff $l_{\textrm{max}} = 20$ as calculated within (a) the hydrodynamic model and (b) the LRA. The bulk, Begrenzung and external terms are given by Eqs.~\eqref{eq:Gammabulk}-\eqref{eq:Gammaext} in panel (a) and Eqs.~\eqref{eq:Gammabulk_local}-\eqref{eq:Gammaext_local} in panel (b). The insets show the full scale of the spectra. Vertical dashed gray lines mark the local dipolar plasmon frequency, $\omega_{\textrm{DP}} = \omega_{p}/\sqrt{3}$, the local surface plasmon frequency, $\omega_{\textrm{SP}} = \omega_{p}/\sqrt{2}$, and the plasma frequency, $\omega_{p}$. (c)-(d) Logarithmic plots of the data shown in panels (a) and (b), respectively. The Begrenzung terms are plotted as the full Begrenzung contribution, $\Gamma^{\textrm{Begr}}$, in a single curve (positive values are plotted as a green curve and negative values as an ochre curve, which is also highlighted by the labels at the bottom of the figure). The insets in panel (c) show the induced charge densities corresponding to the main LSP and CBP excitations for the axial trajectory, $(l,n) = (2,0)$ and $(0,1)$, respectively. Dashed red line represents the contribution of the $\Gamma_{l=2,\,m=0}^{\textrm{ext}}$ term, and the peaks corresponding to the LSP and CBPs in this term are marked with red dots.}
\label{fig:4}
\end{figure*}

To illustrate the derived expressions, we show in Fig.~\ref{fig:4} the EEL spectra calculated for sodium nanoparticles of radius $a = 1.5 \textrm{ nm}$ probed by electron beams with kinetic energy $\mathcal{E}_{k} = 100 \textrm{ keV}$ that pass through the center of the nanoparticle ($b = 0 \textrm{ nm}$).
We plot the EEL spectra calculated within the hydrodynamic model and the LRA in Figs.~\hyperref[fig:4]{\ref*{fig:4}(a)} and \hyperref[fig:4]{\ref*{fig:4}(b)}, respectively, as shaded gray lines, together with different contributions: the bulk term (blue line), the inner and outer Begrenzung terms (green lines) and external term (red line), as given by Eqs.~\eqref{eq:Gammabulk}-\eqref{eq:Gammaext} for the former and Eqs.~\eqref{eq:Gammabulk_local}-\eqref{eq:Gammaext_local} for the latter.
Notice that in Figs.~\hyperref[fig:4]{\ref*{fig:4}(a)} and \hyperref[fig:4]{\ref*{fig:4}(b)} the data are plotted on a linear scale and exceed the plot frame (see insets for uncropped full data).

The low-energy region of the spectra in Fig.~\hyperref[fig:4]{\ref*{fig:4}(a)}, \hbox{$\omega < \omega_{p}$}, is dominated by several LSP excitations whose intensity gradually decreases with increasing energy.
These LSP modes appear blueshifted with respect to the LRA predictions [Fig.~\hyperref[fig:4]{\ref*{fig:4}(b)}] and are not bounded above by the local surface plasmon energy, $\omega_{\textrm{SP}} =\omega_{p}/\sqrt{2}$.
Indeed, within the LRA the LSPs are restricted to the interval $(\omega_{\textrm{DP}},\omega_{\textrm{SP}})$, with $\omega_{\textrm{DP}} =\omega_{p}/\sqrt{3}$ the local dipolar plasmon frequency.
Within the hydrodynamic model [Fig.~\hyperref[fig:4]{\ref*{fig:4}(a)}], multiple CBPs are also excited beyond $\omega_{p}$, and, as a result, a broadened bulk plasmon peak --bulk plasmon envelope-- emerges blueshifted with respect to $\omega_{p}$.
In contrast, in the LRA [Fig.~\hyperref[fig:4]{\ref*{fig:4}(b)}] a unique prominent peak emerges at $\omega_{p}$ with a much higher intensity, as compared to the CBP peaks observed in Fig.~\hyperref[fig:4]{\ref*{fig:4}(a)}.

In Figs.~\hyperref[fig:4]{\ref*{fig:4}(c)} and \hyperref[fig:4]{\ref*{fig:4}(d)} we plot the absolute value of the terms on logarithmic scale to reveal subtle features hidden in Figs.~\hyperref[fig:4]{\ref*{fig:4}(a)} and \hyperref[fig:4]{\ref*{fig:4}(b)}.
The unbound bulk plasmon term (blue line), $\Gamma^{\textrm{bulk}}$, corresponds to the loss produced when the probe perturbs an unbound electron cloud.
This term is centered at $\omega_p$ and is identical in both models.
In the hydrodynamic model, the peak at $\omega_p$ is completely balanced by the Begrenzung term, $\Gamma^{\textrm{Begr}}$, shown as green and ochre lines in Fig.~\hyperref[fig:4]{\ref*{fig:4}(c)}, where we distinguish the $\omega$ range in which the contribution is negative.
As a consequence, only a few modes are required to achieve convergence of the total EEL probability.
In contrast, in the LRA the Begrenzung term, $\Gamma^{\textrm{Begr}}$, does not fully compensate the peak at $\omega_{p}$ [Fig.~\hyperref[fig:4]{\ref*{fig:4}(d)}].
Therefore, the total EEL probability diverges for increasing $l_{\textrm{max}}$ (see discussion about momentum cutoff in Sec. \ref{sec:LRA}).
The Begrenzung term $\Gamma^{\textrm{Begr}}$ can be decomposed into two contributions, $\Gamma^{\textrm{Begr,i}}$ and $\Gamma^{\textrm{Begr,o}}$ (see Fig.~\ref{fig:2} and Sec.~\ref{sec:eloss} for details).
The inner one, $\Gamma^{\textrm{Begr,i}}$, dominates the EEL probability for $\omega > \omega_{p}$ and is primarily responsible for the Begrenzung effect and the excitation of CBPs.
In contrast, the outer contribution, $\Gamma^{\textrm{Begr,o}}$, although nonzero, is much weaker than $\Gamma^{\textrm{Begr,i}}$, in the $\omega > \omega_{p}$ region of Fig.~\hyperref[fig:4]{\ref*{fig:4}(a)}.
Indeed, it mainly contributes to the LSPs, which are shifted to higher frequencies due to the nonlocal dispersion.
These observations have two implications.
First, an efficient excitation of CBPs requires the probe to penetrate inside the nanoparticle.
Second, the probing electron must also remain inside the nanoparticle to interact with the CBPs and experience the induced potential and the associated energy loss.

The distinct nature of LSPs and CBPs is clearly reflected in the cross-sectional maps of their corresponding charge density distributions, shown in the insets of Fig.~\hyperref[fig:4]{\ref*{fig:4}(c)} for modes $(l,n) = (2,0)$ and $(0,1)$, respectively.
For the axial trajectory, only $m=0$ modes contribute, as detailed in the next section.
Notice that within the LRA, the surface charge density associated with the LSPs is assumed to be localized to an infinitely thin layer at the surface.
In contrast, in the hydrodynamic model, the finite compressibility of the electron gas forces this surface charge to spread to the inside of the nanoparticle volume.
As a consequence, the charge density associated with LSPs decays exponentially into the interior of the nanoparticle [see Fig.~\hyperref[fig:3]{\ref*{fig:3}(b)} and the inset of Fig.~\hyperref[fig:4]{\ref*{fig:4}(c)}].

Likewise, the external term $\Gamma^{\textrm{ext}}$ mainly contributes to the excitation of LSPs, whereas its role in exciting CBPs is comparatively minor, being weaker by several orders of magnitude.
The latter is consistent with previous studies in the context of CBP excitation with light and external electron beam trajectories, such as the study by Christensen \textit{et al.} \cite{Christensen_2014_ACSNano}.
Such a contrast between LSPs and CBPs is illustrated by the term $\Gamma_{l=2, \, m=0}^{\textrm{ext}}$, plotted in Fig.~\hyperref[fig:4]{\ref*{fig:4}(c)} as a dashed red line, which shows the dominant excitation of the quadrupolar LSP $(l,n)=(2,0)$ and a much weaker coupling to the CBPs $(2,n>1)$.
Since penetrating trajectories have scarcely been examined in the literature, these weak CBP peaks --also observed for grazing external trajectories-- have been interpreted as the only fingerprint of the nonlocal response of the electron gas at the nanoscale \cite{Christensen_2014_ACSNano}.

Furthermore, the analytical expression of the EEL probability enables a robust multipole-resolved analysis of the spectra, which ultimately depends on two categories of parameters.
(1) Sample-related parameters, such as the size $a$, and the material properties (Wigner-Seitz radius $r_{s}$,  plasma frequency $\omega_{p}$, dispersion parameter $\beta$, and damping $\gamma$); and (2) probe-related parameters, such as the electron impact parameter $b$, and velocity $v$.
In the following sections, we compute and analyze how the EEL probability varies with those parameters that are most frequently explored in experiments: the electron probe's impact parameter and velocity, and the nanoparticle size.

\subsection{Impact parameter dependence}

\begin{figure*}[t!]
 \centering
    \includegraphics[width=1\textwidth]{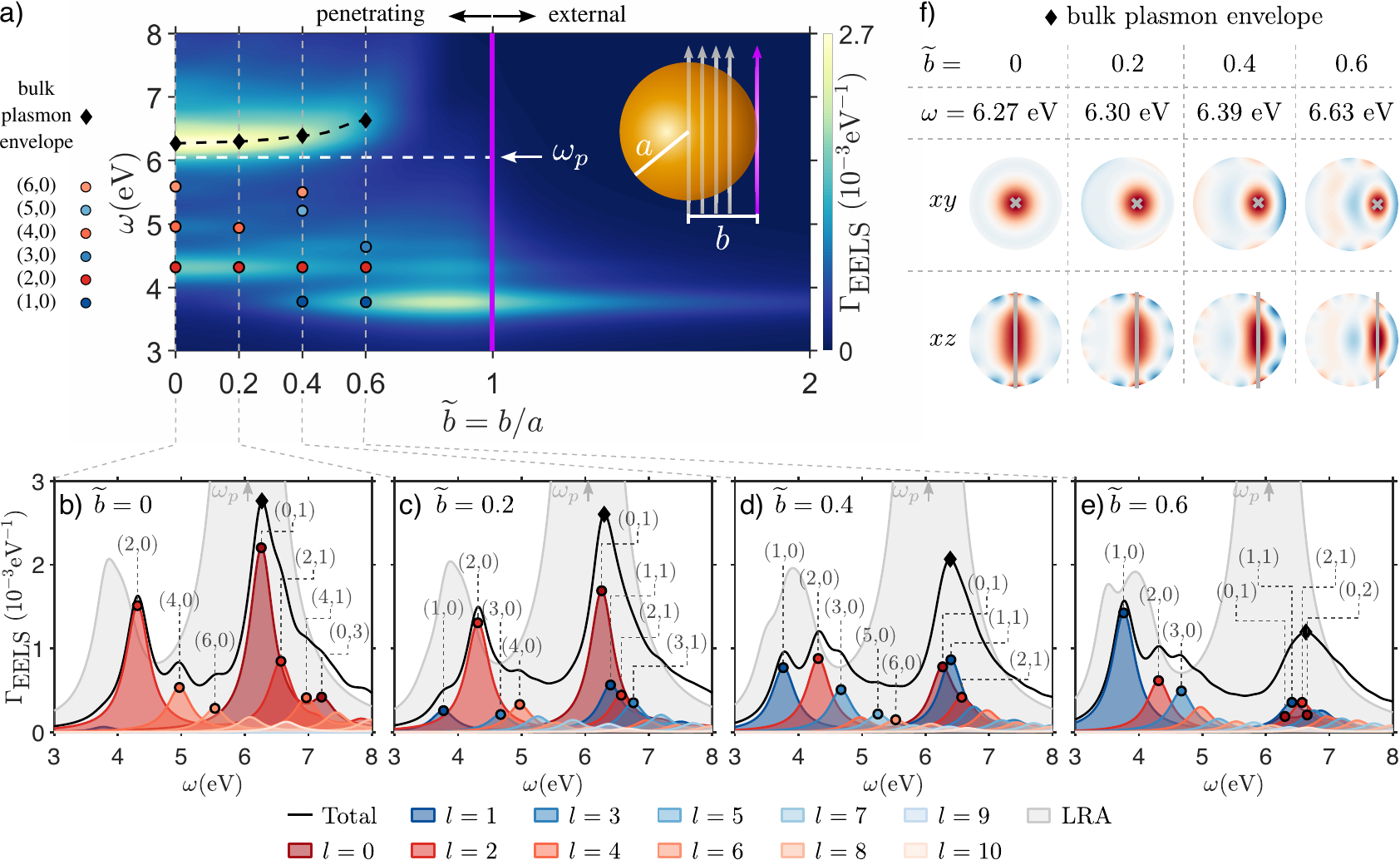}
    \caption{(a) Electron energy loss probability, $\Gamma_{\textrm{EELS}}$, calculated within the hydrodynamic model for a spherical sodium nanoparticle ($r_\textrm{s} = 2.08 \textrm{ \AA}$, $\omega_{p} = 6.05 \textrm{ eV}$) of radius $a=1.5 \textrm{ nm}$, as a function of the energy lost by the electron probe, $\omega$, and the reduced impact parameter, $\widetilde{b}=b/a$. The kinetic energy of the electron beam is $\mathcal{E}_{k} = 100 \textrm{ keV}$. The vertical purple straight line highlights the grazing trajectories ($b=a$) and the horizontal dashed white line marks the plasma frequency $\omega_{p}$. The dots in colors point the positions of the LSP peaks, and the black diamonds point out the bulk plasmon envelope for spectra with reduced impact parameters $\widetilde{b} = 0, 0.2, 0.4 \textrm{ and } 0.6$ (vertical dashed gray lines). (b)-(e) Electron energy loss probability (black line) and the partial contributions from $l=0-10$ modes (we include the contribution from all $m$ for each $l$) for normalized impact parameters $\widetilde{b} = 0, 0.2, 0.4 \textrm{ and } 0.6$. The dominant modes are highlighted with dots and labeled as $(l,n)$. The gray shaded curve represents the electron energy loss probability calculated within the LRA. (f) Induced charge density maps for selected energies $\omega$ corresponding to the position of the bulk plasmon envelope.
    }
\label{fig:5}
\end{figure*}

To investigate the dependence of the EEL probability on the impact parameter, we consider the same model sodium nanoparticle used in the previous section.
Figure \hyperref[fig:5]{\ref*{fig:5}(a)} shows the EEL probability calculated as a function of the energy loss $\hbar \omega$, and of the reduced impact parameter $\widetilde{b}=b/a$, with $a$ the radius of the nanoparticle, for both internal ($\widetilde{b}<1$) and external ($\widetilde{b}>1$) trajectories of the probing electron.

For external trajectories and large impact parameters $\widetilde{b} \sim 2$ [right-hand side of Fig. \hyperref[fig:5]{\ref*{fig:5}(a)}], the dipolar LSP mode $(l,n)=(1,0)$ dominates the spectra.
As the impact parameter decreases, higher-order LSP modes are excited, especially at grazing trajectories ($\widetilde{b} \sim 1$) \cite{Ferrell_1985_PhysRevLett, Ferrell_1987_PhysRevB}.
Noticeably, there is no trace of CBP excitation for external trajectories.
Due to their weak coupling to external electron beams, CBPs are only distinguishable in the spectra when a logarithmic scale is used \cite{Christensen_2014_ACSNano}.
However, this is not the case for penetrating trajectories [left-hand side of Fig. \hyperref[fig:5]{\ref*{fig:5}(a)}], which exhibit much more complex spectra with a stronger dependence on the impact parameter.
The $(1,0)$ LSP mode is the dominant mode excited for beam trajectories passing through the outer half of the nanoparticle $(0.5 < \widetilde{b} < 1)$, but it fades away for trajectories crossing the inner half ($\widetilde{b} < 0.5$).
The decay of this mode is balanced by an increase in the intensity of the quadrupolar LSP mode $(2,0)$, located at $\sim 4.3 \textrm{ eV}$.

Nevertheless, the dominant feature for penetrating trajectories in Fig.~\hyperref[fig:5]{\ref*{fig:5}(a)} is the emergence of CBPs as a single peak --referred to as the bulk plasmon envelope-- above $6 \textrm{ eV}$, which exhibits two remarkable trends.
First, the bulk plasmon envelope does not appear immediately for $\widetilde{b}=1$ but emerges for smaller impact parameters, a trend also observed in \textit{ab initio} calculations for nonspherical atomistic nanoclusters \cite{Urbieta_2024_PhysChemChemPhys} and within nonlocal models for planar interfaces \cite{Zabala_1990_Ultramicroscopy}.
This behavior contrasts with predictions of the LRA, which substantially overestimates bulk plasmon losses for penetrating trajectories grazing the nanoparticle surface.
Second, the excitation of new CBPs produces an effective blueshift of the bulk plasmon envelope with increasing impact parameter, a feature not captured by the LRA.
Later in this work, we will explore both of these trends as a function of nanoparticle size.

To illustrate the underlying modal structure of the peaks observed in Fig.~\hyperref[fig:5]{\ref*{fig:5}(a)}, we plot four selected spectra for the reduced impact parameters $\widetilde{b} = 0, 0.2, 0.4 \textrm{ and } 0.6$ in Figs.~\hyperref[fig:5]{\ref*{fig:5}(b)}-\hyperref[fig:5]{\ref*{fig:5}(e)}.
The total EEL probability calculated with Eq.~\eqref{eq:EELprobabilitygen} is shown in black, and the contribution of each $l$ mode is plotted separately in red for even $l$ and in blue for odd $l$.
For each trajectory, the main contributing modes are highlighted with dots and labeled as $(l,n)$ correspondingly.
As a reference, we also show the EEL probability obtained within the LRA for a sphere with identical parameters (shaded gray area) \cite{Rivacoba_1990_ScannMicrosc, Rivacoba_2000_ProgrSurfSci}.

The prevalence of even-$l$ modes over the odd-$l$ modes for the axial trajectory ($\widetilde{b} = 0$) is evident from Fig.~\hyperref[fig:5]{\ref*{fig:5}(b)}, and the progressive activation of odd-$l$ modes as the impact parameter increases can be traced in Figs.~\hyperref[fig:5]{\ref*{fig:5}(c)}-\hyperref[fig:5]{\ref*{fig:5}(e)} for both LSP and CBP modes.
Notably, the activation of odd-$l$ CBP modes is responsible for the effective blueshift of the bulk plasmon envelope observed for increasing impact parameters, as reflected in the induced charge density distributions shown in Fig.~\hyperref[fig:5]{\ref*{fig:5}(f)}.

The induced charge densities in the $xy$ and $xz$ cross sections of the nanoparticle at the energy corresponding to the bulk plasmon envelope [black diamond on the total EEL spectrum (black line) in Figs.~\hyperref[fig:5]{\ref*{fig:5}(b)}-\hyperref[fig:5]{\ref*{fig:5}(e)}] correspond to the electron trajectories from the center ($b=0$) outward, toward the surface $b\rightarrow a$.
It is evident that the azimuthal symmetry associated with the $b=0$ trajectory is broken when $b \neq 0$.
As a consequence, multiple CBP modes are excited for each impact parameter, overlapping in energy with specific LSPs, which manifests as ripples in the induced charge density near the nanoparticle surface.

Similarly to LRA predictions \cite{Rivacoba_2000_ProgrSurfSci}, within the hydrodynamic model the activation and deactivation of LSPs --and to some extent of CBPs-- is ruled by two parameters: (1) the electron beam's traveling time inside the nanoparticle and (2) the parity of the $l$ number associated with each mode.
This traveling time is given by $\Delta t=L/v$, where $L=2\sqrt{a^2-b^2}$ is the path length.
In the $\omega$ domain, such time lapse translates into a phase difference of the $\omega$ component of the electric field generated by the electron beam at the entry and exit points of the nanoparticle, given by $\Delta \varphi^{\textrm{ext}} (b) =  2 \omega \sqrt{a^2-b^2}/v$.
For instance, the excitation of a given LSP mode is maximized when the phase difference of the associated induced charge density at the probe's entry and exit points, $\Delta \varphi(b)$, matches the phase difference of the field created by the electron beam, $\Delta \varphi^{\textrm{ext}} (b)$.
For a central trajectory ($b=0$), the phase difference for the $m=0$ dipolar and quadrupolar LSP modes are $\Delta \varphi(0) = \pi$ and $\Delta \varphi(0) = 0$, respectively.
Indeed, for $m=0$, odd-$l$ and even-$l$ modes exhibit phase differences $\Delta \varphi(0) = \pi$ and $\Delta \varphi(0) = 0$, respectively.
For a swift 100 keV electron beam traversing the nanoparticle through its center, the time elapsed between the entry and exit points is so small that the corresponding phase difference is negligible for the frequencies corresponding to the LSPs, i.e., $\Delta \varphi^{\textrm{ext}} (0) \rightarrow 0$.
The phase-matching condition then selects even-$l$ modes for axial trajectories.
This selection rule explains the variations in the intensity of the LSPs and most CBPs as a function of the impact parameter observed in Figs.~\hyperref[fig:5]{\ref*{fig:5}(a)}-\hyperref[fig:5]{\ref*{fig:5}(e)}, and it is illustrated in Fig.~\hyperref[fig:5]{\ref*{fig:5}(f)}, where the induced charge densities at the beam’s entry and exit points are nearly in phase ($\Delta \varphi \simeq 0$).

Nevertheless, beyond the relationship between the multipole number ($l$ parity) and the impact parameter of the electron beam, there is a subtler underlying trend in these spectra.
In Fig.~\hyperref[fig:5]{\ref*{fig:5}(b)}, for $\widetilde{b}=0$, in addition to the main mode $(0,1)$, we observe a secondary peak corresponding to the $(0,3)$ mode, while notably the $(0,2)$ mode is missing.
Upon inspection, this trend also holds for higher-order CBPs: modes with even $n$ radial numbers, $(0,2k)$ with $k \in \mathbb{N}$, are much weaker, while odd-$n$ modes, $(0,2k+1)$, are excited, although their EEL probability decreases monotonically with increasing $n$.

This behavior is observed not only for the $l=0$ monopole but also for other even-$l$ modes, in which even-$n$ modes are suppressed and odd-$n$ modes appear.
Conversely, both the $(0,1)$ and $(0,2)$ CBP modes are excited for $\widetilde{b}=0.6$ [see Fig.~\hyperref[fig:5]{\ref*{fig:5}(e)}], implying the existence of a selection rule related to the parity of $n$ (recall that $n$ is directly linked to the number of radial oscillations of the CBP mode, as discussed in Sec. \ref{sec:theory_HM}).
We analyze this relationship in the following section.

\subsection{Velocity dependent excitation of CBPs}

\begin{figure*}[t!]
 \centering
    \includegraphics[width=1\textwidth]{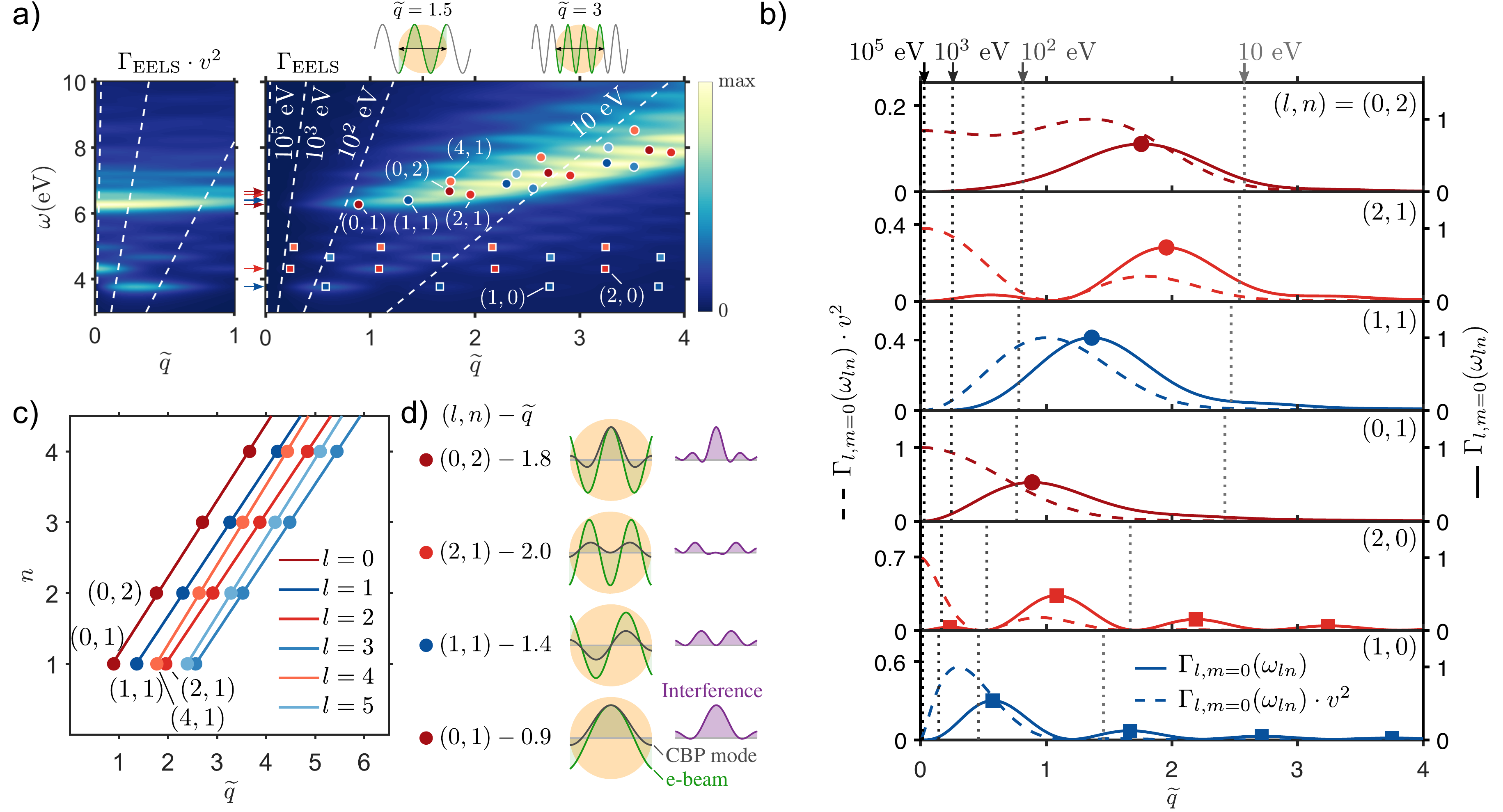}
    \caption{(a) Electron energy loss probability maps, calculated within the hydrodynamic model, as a function of the energy loss, $\omega$, and reduced wave number, $\widetilde{q}=\frac{\omega a}{v \pi}$, of the excitation field within a nanoparticle of radius $a=1.5 \textrm{ nm}$ for central electron trajectories ($b=0$). Dashed white lines show spectral cuts corresponding to electron beams with energies $\mathcal{E}_{k}=10^{5}, 10^{3}, 10^{2} \textrm{ and } 10 \textrm{ eV}$. Squares highlight the peaks corresponding to LSPs. Circles correspond to the positions at which the contribution of each CBP is maximum. (b) Solid line (right axis): contribution, $\Gamma_{l,m=0}$, from LSPs $(1,0)$ and $(2,0)$, and CBPs $(0,1)$, $(1,1)$, $(2,1)$ and $(0,2)$ to the EEL probability as a function of $\widetilde{q}$ at their respective resonance energy, $\omega = \omega_{ln}$, highlighted with arrows on the axis of $\omega$ in panel (a). All values are normalized to the maximum contribution of $(1,1)$. Dashed line (left axis): same data multiplied by $v^2$. All values are normalized to the maximum contribution of $(0,1)$.
    Vertical dashed gray lines highlight the $\widetilde{q}$ corresponding to electron beam energies $\mathcal{E}_{k}=10^{5}, 10^{3}, 10^{2} \textrm{ and } 10 \textrm{ eV}$.
    (c) Values of $\widetilde{q}$ that maximize the contribution from each CBP. (d) Sketches of the charge density (gray) associated with each CBP along the electron beam path, together with the $\omega$-component electric field associated with the electron beam (green) for the corresponding $\widetilde{q}$, highlighted in panel (c), and the interference (purple) between both of them.
    The nanoparticle is outlined in light orange.
    }
\label{fig:6}
\end{figure*}

To identify the underlying mechanism governing the activation of CBP modes and its dependence on the parity of the radial number $n$, we have calculated the EEL probability for the central trajectory as a function of the electron beam velocity.
In the frequency domain, the number of oscillations of the $\omega$ component of the electric field generated by the electron beam inside the nanoparticle is determined by the nanoparticle radius $a$, the impact parameter $b$, and the electron beam's velocity $v$.
For a central trajectory ($b=0$), the number of oscillations along the electron path inside the nanoparticle reduces to $\widetilde{q} = \frac{\omega a}{\pi v}$, which we refer to as the reduced wave number (or momentum) [see top insets in Fig.~\hyperref[fig:6]{\ref*{fig:6}(a)}].
Thus, by varying $a$ or $v$, one can tune the value of $\widetilde{q}$ for each $\omega$.
In what follows, we simplify the analysis by keeping $a$ constant, therefore fixing the mode energies $\omega_{ln}$, so that the values of $\widetilde{q}$ are tuned solely through $v$.

These trends are shown in Fig.~\hyperref[fig:6]{\ref*{fig:6}(a)} for a nanoparticle with radius $1.5 \textrm{ nm}$, where $\Gamma_{\textrm{EELS}}$ is displayed as a function of $\widetilde{q}$ and $\omega$.
In the left panel the loss probability is multiplied by $v^2$ to highlight the behavior at small $\widetilde{q}$.
As a reference, we also include spectral cuts corresponding to electron beams with fixed kinetic energies $\mathcal{E}_{k}=10^{5}, 10^{3}, 10^{2}, \textrm{ and } 10 \textrm{ eV}$ indicated with dashed lines.
The figure further displays the partial contributions of the $(l,n)$ modes to the EEL spectra.
As expected, a periodic activation and deactivation of odd- and even-$l$ modes is observed for LSPs as a function of $\widetilde{q}$, highlighted by blue and red squares, marking the local maxima of the LSPs $(1,0)$, $(2,0)$, $(3,0)$ and $(4,0)$.
Specifically, values $\widetilde{q} = \frac{1}{2}+k$, with $k \in \mathbb{N}$, yield phase differences  $\Delta \varphi^{\textrm{ext}}(0) = (1+2k) \pi$, which predominantly excite odd-$l$ LSP modes.
Conversely, integer values of $\widetilde{q}$ give $\Delta \varphi^{\textrm{ext}}(0) = 2k \pi$, favoring the excitation of even-$l$ LSP modes \cite{Rivacoba_2000_ProgrSurfSci}.
On the other hand, the intensity of higher-order CBP modes increases with increasing $\widetilde{q}$, at the expense of a reduction in the intensity in lower-order CBP modes, leading to an effective blueshift of the bulk plasmon envelope in the spectra.
This dependence on $\widetilde{q}$ for each $(l,n)$ mode is explored in more detail in Fig. \hyperref[fig:6]{\ref*{fig:6}(b)}, where the contribution to the EEL probability, $\Gamma_{l,m=0}$, is plotted as a function of $\widetilde{q}$ for the LSPs $(1,0)$ and $(2,0)$, and the CBPs $(0,1)$, $(1,1)$, $(2,1)$ and $(0,2)$, at their respective resonance energies, $\omega_{ln}$.
To analyze the spectra for fast electron beams ($\mathcal{E}_{k} > 10 \textrm{ keV}$), the same data multiplied by $v^2$, $\Gamma_{l,m=0} v^2$, are shown in Fig.~\hyperref[fig:6]{\ref*{fig:6}(b)} as dashed lines.
These results demonstrate that while penetrating trajectories through the center of small nanoparticles predominantly excite the $(0,1)$ CBP, higher-order CBPs can be activated for appropriately chosen parameters.
More importantly, each $(l,n)$ mode exhibits an optimal value of $\widetilde{q}$ that maximizes its excitation, marked by dots in Figs.~\hyperref[fig:6]{\ref*{fig:6}(a)} and \hyperref[fig:6]{\ref*{fig:6}(b)}.
The full set of correlations between $\widetilde{q}$ and the $(l,n)$ indices of the CBP modes is summarized in Fig.~\hyperref[fig:6]{\ref*{fig:6}(c)}.

Direct inspection of Eq.~\eqref{eq:GammaBegr_i} shows that the dominant contribution to the excitation of the CBP modes depends primarily on the terms involving the integrals $\mathcal{I}_{lm}$ and, in particular, $\mathcal{J}_{lm}$ [see Eqs.~\eqref{eq:Ilm} and \eqref{eq:Jlm}], which define the coupling of the CBP modes to the electron beam.
It is instructive to focus on the latter term for the central trajectory.
In this case, $\mathcal{J}_{lm}$ reduces to:
\begin{align}
    \mathcal{J}_{l0}(0,a) = \mu \int_{0}^{a} dz \, j_l(\mu z) g_{l0} (\omega z / v),
\end{align}
which accounts for the interference between the CBP modes, described by $j_{l}(\mu r)$ [see Eq.~\eqref{eq:chargedensityhomogeneous}], and the electron beam, given by $g_{lm}(\omega z / v)$ [see Eqs.~\eqref{eq:extchargesph}-\eqref{eq:glm}], along its trajectory.

\begin{figure*}[t!]
 \centering
    \includegraphics[width=1\textwidth]{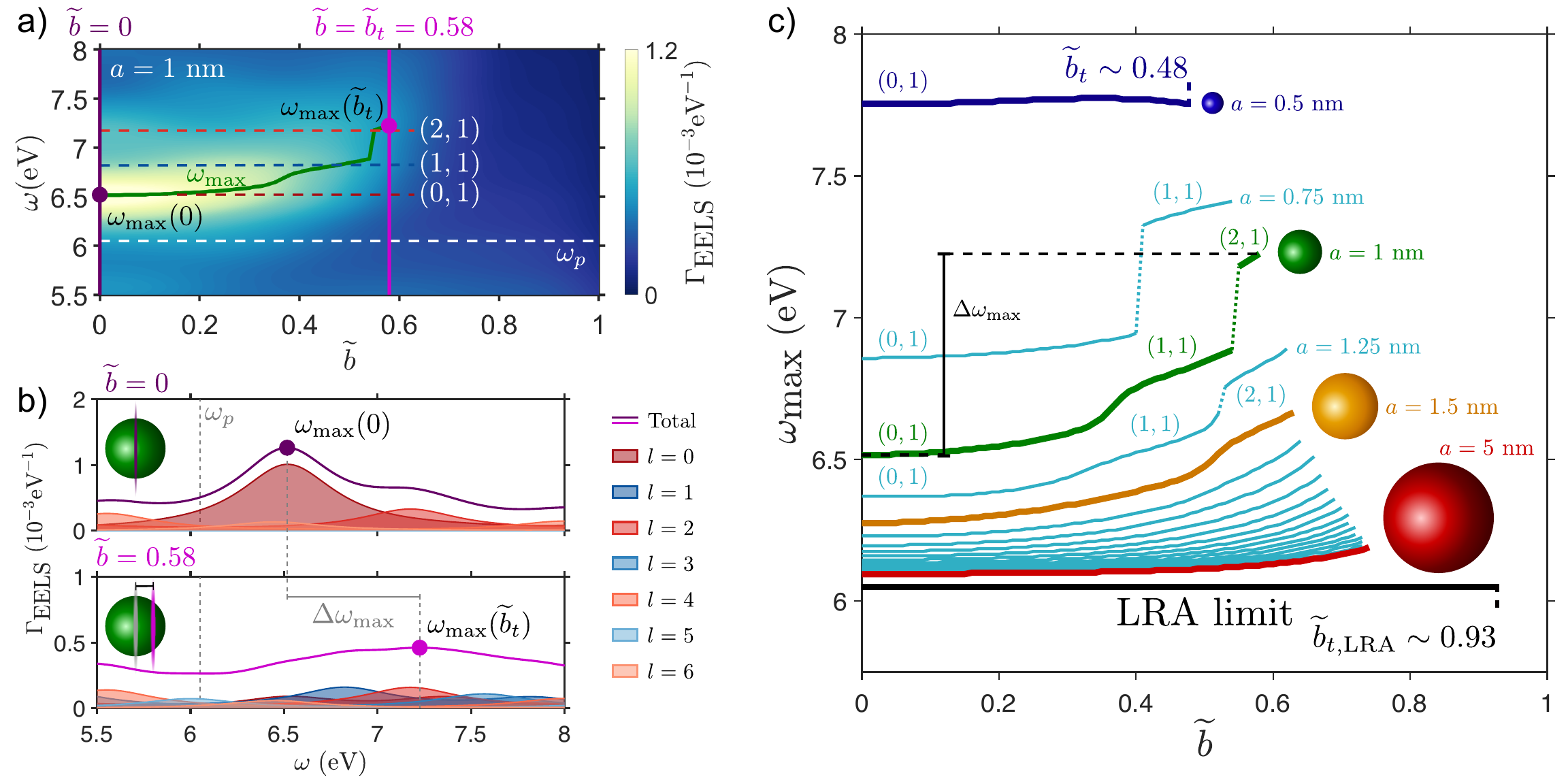}
    \caption{(a) Electron energy loss probability as a function of the reduced impact parameter, $\widetilde{b} = b/a$, for a nanoparticle of radius $a=1 \textrm{ nm}$ probed by an electron beam with energy $\mathcal{E}_{k} = 100 \textrm{ keV}$.
    The solid green line follows the energy of the dominant CBP peak for each impact parameter, from $\widetilde{b}=0$ to $\widetilde{b}_{t}$.
    The positions of the local plasma frequency, $\omega_{p}$, and CBP modes $(0,1)$, $(1,1)$ and $(2,1)$ are plotted with dashed lines as a reference.
    (b) EEL spectra and contribution of the $l=0-10$ modes for $\widetilde{b}=0$ (top panel) and $\widetilde{b}=\widetilde{b}_{t}$ (bottom panel), respectively.
    c) Frequency of the dominant CBP peak, $\omega_{\textrm{max}}$, as a function of $\widetilde{b}$ for $a=0.5 - 5 \textrm{ nm}$.
    The horizontal black line marks the results obtained within the LRA.
    The solid green curve corresponds to $a = 1 \textrm{ nm}$ [solid green line in panel (a)], and the yellow one to $a = 1.5 \textrm{ nm}$ [plotted as a dashed black line in Fig.~\hyperref[fig:5]{\ref*{fig:5}(a)}].
    }
\label{fig:7}
\end{figure*}

Constructive interference leads to an increase in the EEL probability, as sketched in Fig.~\hyperref[fig:6]{\ref*{fig:6}(d)} for the central trajectory ($\widetilde{b}=0$).
The charge density associated with the CBP modes (gray) for $(0,1)$, $(0,2)$, $(1,1)$ and $(2,1)$ [also represented in Fig.~\hyperref[fig:3]{\ref*{fig:3}(b)}, proportional to $j_l(\mu r)$], and the excitation field (green) at 
$\widetilde{q} \sim 0.9, 1.8, 1.4 \textrm{ and } 2$ [proportional to $g_{l0}(\omega z/v)$], which correspond to peaks in the spectrum, produce the interference patterns (purple shade) shown in the right of Fig.~\hyperref[fig:6]{\ref*{fig:6}(d)}.
As a rule of thumb, Fig.~\hyperref[fig:6]{\ref*{fig:6}(c)} suggests that $\widetilde{q}= \frac{l}{2} + n$ for the lowest $l$ modes, although a mismatch already appears for $n=1$, attributed to the deviation of spherical Bessel functions from a pure harmonic oscillation.
The interplay between $\widetilde{q}$ and $(l,n)$ indices reveals that CBP excitation follows a more subtle selection rule, depending on the full trajectory, as implied by Eqs. \eqref{eq:Ilm}-\eqref{eq:Ylm}.
In contrast, LSP excitation depends mainly on the phase matching between the entry and exit points.

This explains the absence of a peak corresponding to $(0,2)$ in Fig.~\hyperref[fig:5]{\ref*{fig:5}(b)}, due to the nearly complete destructive interference between the electron beam and the $(0,2)$ mode for fast electron beams and small nanoparticles ($\widetilde{q} \rightarrow 0$).
Notice that the nonrecoil approximation, which assumes a constant electron beam velocity, fails at low kinetic energies (below 1 keV), a regime beyond the scope of this work.

\subsection{Size effects in the excitation of CBPs}

Once the selection rules governing the excitation of the LSP and CBP modes have been established, we turn our attention to the blueshift of the bulk plasmon envelope with increasing impact parameters observed in Fig.~\ref{fig:5}.
As shown there, the bulk plasmon envelope decays rapidly for $\widetilde{b} > 0.6$.
We refer to this threshold as the bulk plasmon activation threshold: the impact parameter $\widetilde{b}_{t}$ for which the peak intensity of the bulk plasmon envelope reaches $1/e \sim 0.37$ of its maximum value at $\widetilde{b} = 0$.
As discussed above, this trend is in clear contrast with the LRA description, where the intensity of the bulk plasmon peak is proportional to the length of the path traveled by the electron inside the nanoparticle [see Eq.~\eqref{eq:Gamma_bulk_local_log}].
Within the LRA, the dependence on the impact parameter is given by $\Gamma_{\textrm{LRA}}^{\textrm{bulk}} \propto (1 - \widetilde{b}^2)^{1/2}$, which sets the bulk plasmon activation threshold at $\widetilde{b}_{t,\textrm{LRA}} \sim 0.93$.
Indeed, EELS experiments on sub-10-nm nanoparticles \cite{Raza_2015_NatComm, Hobbs_2016_NanoLett} have revealed that an efficient excitation of the bulk plasmon requires smaller impact parameters than those predicted by the LRA.
With this discrepancy in mind, we explore in this section how the nanoparticle size influences the EEL spectra as a function of the impact parameter, paying special attention to the blueshift and the variation of the CBP peak intensity with increasing impact parameters.

In Fig.~\hyperref[fig:7]{\ref*{fig:7}(a)} we show a close-up of the EEL spectra around $\omega_{p}$ as a function of the reduced impact parameter $\widetilde{b}$ for a nanoparticle of radius $a=1 \textrm{ nm}$, probed by a $100 \textrm{ keV}$ electron beam.
The solid green line traces the energy of the most intense CBP peak for each impact parameter, $\omega_{\textrm{max}}(\widetilde{b})$ (thereafter, we omit the explicit dependence on $\widetilde{b}$ unless necessary).
The purple dot at \hbox{$\widetilde{b} = 0$ and $\omega = 6.5 \textrm{ eV}$} marks the position of the most intense CBP peak for the central trajectory, whereas the pink dot at $\widetilde{b} = \widetilde{b}_{t}\simeq 0.58$ and $\omega = 7.2 \textrm{ eV}$ highlights the most intense CBP peak at the bulk plasmon activation threshold.
For reference, the local plasma frequency ($\omega_{p} = 6.05 \textrm{ eV}$) and CBP modes $(0,1)$, $(1,1)$ and $(2,1)$ are indicated with dashed lines as a guide to the eye.

The spectra corresponding to the two highlighted impact parameters are displayed in the top and bottom panels in Fig.~\hyperref[fig:7]{\ref*{fig:7}(b)}, together with the partial contributions of the $l=0-6$ modes in each case (red for even-$l$ and blue for odd-$l$ modes).
These results demonstrate  that the characteristic blueshift observed for increasing impact parameters arises from the relative enhancement in the excitation of higher-order CBP modes.
However, when comparing these spectra with those of the larger nanoparticle of $a=1.5\textrm{ nm}$ at $b=0$ in Fig.~\hyperref[fig:5]{\ref*{fig:5}(c)}, one notices that the energies of the CBP modes are systematically blueshifted for the smaller nanoparticle.
More interestingly, the blueshift of the bulk plasmon envelope is no longer smooth: abrupt jumps emerge in $\omega_{\textrm{max}}$ as the impact parameter increases [see the solid green line in Fig.~\hyperref[fig:7]{\ref*{fig:7}(a)}].
These jumps can be traced to specific $(l,n)$ as displayed in Fig.~\hyperref[fig:7]{\ref*{fig:7}(c)}.
Moreover, the activation threshold $\widetilde{b}_{t}$ decreases as $a$ decreases, as illustrated in Fig.~\hyperref[fig:7]{\ref*{fig:7}(c)}, where we plot the energy of the most intense CBP peak, $\omega_{\textrm{max}}$, as a function of the reduced impact parameter $\widetilde{b}$ for nanoparticle radii in the range $a=0.5-5 \textrm{ nm}$.

Figure \hyperref[fig:7]{\ref*{fig:7}(c)} summarizes the trends discussed above.
First, $\omega_{\textrm{max}}(0)$ increases as $a$ decreases [see the labels on the right-hand side of Fig.~\hyperref[fig:7]{\ref*{fig:7}(c)}], showing the characteristic blueshift with decreasing radii due to dispersion.
Second, for a fixed radius $a$, a blueshift with increasing impact parameter is observed, arising from the excitation of higher-order CBP modes (see, for instance, the yellow curve).
Third, the threshold impact parameter $\widetilde{b}_{t}$ decreases with decreasing radius (compare the lengths of the red and the yellow curves).
Moreover, the energy shift between the position of the bulk plasmon envelope at $\widetilde{b}=0$ and at $\widetilde{b} = \widetilde{b}_{t}$, \hbox{$\Delta \omega_{\textrm{max}} = \omega_{\textrm{max}}(\widetilde{b}_{t}) - \omega_{\textrm{max}}(0)$,} also increases as $a$ decreases.
This can be seen by comparing, for instance, the blueshifts of the red ($a = 5 \textrm{ nm}$), yellow ($a = 1.5 \textrm{ nm}$) and green ($a = 1 \textrm{ nm}$) curves.
This increase in the energy shift for decreasing $a$ is due to the stronger dispersion of the higher-order CBP modes at smaller radii.
Indeed, for $a<1.5 \textrm { nm}$, the widening of the energy gaps between CBP modes leads to the abrupt jumps observed in the evolution of $\omega_{\textrm{max}}$, also visible in Fig.~\hyperref[fig:7]{\ref*{fig:7}(a)}.
As a consequence, this widening of the energy gaps also leads, for $a<1 \textrm{ nm}$, to a decrease of $\Delta \omega_{\textrm{max}}$ with decreasing $a$ (see the dark blue curve).
As the radius $a$ increases, the predictions of the hydrodynamic model gradually converge to those of the LRA [see Sec.~\ref{sec:LRA}].
However, the LRA fails completely, not only in predicting the mentioned blueshifts, but also in capturing the dependence of the bulk plasmon activation threshold on $a$, since this threshold parameter remains fixed at $\widetilde{b}_{t,\textrm{LRA}} \sim 0.93$, independently of $a$.

The trends related to the blueshift, as well as the deviation between the local and nonlocal descriptions, can be fully understood by examining the CBP modal structure and its dependence on the nanoparticle size [see Eqs.~\eqref{eq:modecond} and \eqref{eq:modecondzero}].
The size dependence of the bulk plasmon activation threshold may also be rationalized in the following terms: for sufficiently large nanoparticles, the contributions of higher-order CBPs to the total EEL probability overlap with those of the lower-order CBP modes, forming a single peak near $\omega_{p}$, that is, the bulk plasmon envelope.
For decreasing nanoparticle sizes, the increase in the energy gap between the CBPs reduces this overlap to the point where for $a\leq0.5 \textrm{ nm}$ the overlap among the lowest CBP modes becomes negligible, and solely the contribution of the $(0,1)$ mode exceeds the bulk plasmon activation threshold. 
Thus, from a physical standpoint the decrease in $\widetilde{b}_{t}$ with decreasing $a$ can be associated with the finite compressibility of the electron density considered within the hydrodynamic model.

We have focused on spherical nanoparticles because their symmetry allows for a more straightforward analytical treatment. However, breaking this symmetry affects LSPs and CBPs in distinct ways.
For LSPs, deviations from sphericity modify the angular distribution of surface charge and lift the $(l,m,n)$ degeneracy \cite{Collins_2018_PhysRevB}, giving rise to new modes with shifted energies and modified charge patterns.
This splitting can be large enough to generate additional peaks in the EEL spectra, as shown by \textit{ab initio} atomistic time-dependent density function theory (TDDFT) calculations of metallic clusters \cite{Urbieta_2024_PhysChemChemPhys}.

CBPs also lose their spherical degeneracy under shape distortions, and their modes no longer follow the $(l,m,n)$ classification.
Moreover, their characteristic radial length also becomes orientation-dependent.
However, because CBP modes are much more closely spaced in energy compared to LSPs, the appearance of clearly resolvable new peaks is unlikely.
For nanoparticles that slightly depart from a spherical symmetry —for example, typical polyhedral morphologies such as icosahedral particles— we expect very similar results \cite{Candelas_2025_JPhysChemLett}.
Although the bulk plasmon spectral envelope may broaden and its intensity slightly decreases due to the excitation of multiple CBPs at nearby energies, for such geometries the spectra may often be accurately reproduced using the spherical nanoparticle approximation.
In contrast, strongly anisotropic structures, such as nanorods or nanoellipsoids, should exhibit more pronounced CBP energy shifts.

\section{Conclusions}

In conclusion, we have developed a nonlocal analytical expression for the EEL spectra of electron beams probing sub-5-nm spherical nanoparticles within a linear hydrodynamic model.
The derived expressions [Eqs.~\eqref{eq:EELprobabilitygen} and \eqref{eq:Gammabulk}-\eqref{eq:Gammaext}] describe the EEL probability within a multipolar basis, for both penetrating and external electron trajectories.
They shed light on the nature of the Begrenzung effect and its role in the excitation of CBPs for penetrating conditions.
We have applied these expressions to sodium nanoparticles as a canonical example of a free electron gas, demonstrating that, unlike external trajectories, penetrating trajectories efficiently excite both LSP and CBP modes.

Our analysis reveals two main features of the spectral envelope of the CBPs, consistent with \textit{ab initio} calculations and experiments, but not captured by widely used local approaches.
First, efficient activation of the bulk plasmon envelope requires a threshold impact parameter $\widetilde{b}_{t}$.
Second, the bulk plasmon envelope exhibits a pronounced blueshift as the impact parameter shifts from the center of the nanoparticle toward its surface ($b \rightarrow a$), reflecting the high sensitivity of the EEL probability to the electron beam's impact parameter.

We further find that CBP excitation follows subtler selection rules than those governing LSPs, determined both by the path of the electron beam inside the nanoparticle and by the symmetry of the excited modes.
This sensitivity explains the effective blueshift of the bulk plasmon peak observed experimentally in Al disks \cite{Hobbs_2016_NanoLett} and Bi nanoparticles \cite{BorjaUrby_2019_JElectronSpectrosRelatPhenomena} with increasing impact parameters, which we attribute to the preferential excitation of higher-order CBPs by penetrating trajectories near the nanoparticle surface.

Additionally, we show that the excitation of CBPs in EELS also depends on the electron beam kinetic energy.
For axial trajectories, the relative excitation of a given $(l,n)$ CBP mode is maximized at specific values of the reduced wave number $\widetilde{q}$ that correlate with the $(l,n)$ numbers, as shown in Fig.~\hyperref[fig:6]{\ref*{fig:6}(c)}.
Size effects further modulate these trends: as the nanoparticle radius decreases, our analytical model predicts an increasing blueshift of the CBP modes and an increase in the mode dispersion.
The associated increase in the energy spacing of CBPs produces a decrease in the overlap of CBPs, which gives rise to a reduction of the threshold impact parameter $\widetilde{b}_{t}$, enhances the relative blueshift with increasing $b$, and leads to step like shifts in the bulk plasmon envelope with increasing $b$ for sufficiently small nanoparticles ($a< 1.5 \, \textrm{nm}$).

Methodologically, the hydrodynamic model achieves multipolar convergence without the \textit{ad hoc} momentum cutoffs required by the LRA treatments of bulk losses, offering a compact and comprehensive framework to analyze the bulk plasmon range of EEL spectra.
Within the scope of the present work --quasistatic limit, $\varepsilon_\infty=\varepsilon_m=1$, nonrecoil approximation, and spherical symmetry-- these results clearly describe how CBP excitation depends on electron beam parameters and particle size.
In addition, they provide a useful framework for interpreting CBP-sensitive STEM-EELS experiments as well as \textit{ab initio} atomistic TDDFT calculations \cite{Urbieta_2024_PhysChemChemPhys, Candelas_2025_JPhysChemLett}.

\section*{Data Availability Statement}
The data that support the findings of this article are openly available \cite{suppdata}.

\begin{acknowledgments}
We acknowledge financial support through Grant No. PID2022-139579NB-I00, funded by MICIU/AEI/10.13039/501100011033 and by ERDF/EU; and through Grant No. IT1526-22, funded by the Department of Science, Universities and Innovation of the Basque Government.
\end{acknowledgments}

\appendix

\section{Calculation of the induced charge density} \label{app:a}
In this Appendix, we derive the induced charge density inside a spherical nanoparticle excited by a penetrating electron beam.
We solve the inhomogeneous Helmholtz equation [Eq.~\eqref{eq:Helmholtz}] and enforce the hydrodynamic boundary condition [Eq.~\eqref{eq:BC}].
The aim is to express the induced charge density as a multipole expansion suitable for obtaining the induced potential and, ultimately, the EEL probability in a spherical nanoparticle.
For the sake of simplicity, throughout this appendix, we omit explicitly $\omega$ dependence of the terms.

\subsection{Multipole expansion of the external charge density} \label{app:a.1}
Exploiting the spherical symmetry of the system, we expand the external charge density [Eq.~\eqref{eq:extchargedens}] in terms of spherical harmonics, $Y_{l}^{m}(\Omega)$:
\begin{align}
\varrho^{\textrm{ext}}(\textbf{r}) & = \sum_{l=0}^{\infty} \sum_{m=-l}^{l} \varrho_{lm}^{\textrm{ext}}(r) Y_{l}^{m} (\Omega),  \label{eq:chargedensityextgeneral}
\end{align}
where
\begin{align}
\varrho_{lm}^{\textrm{ext}} (r) & = - \frac{2 \, \alpha_{lm}}{v r^2 \sqrt{1 - \frac{b^2}{r^2}}} \; \Theta \bigg{(}\frac{r}{b} - 1 \bigg{)}\Pi_{lm} (r) \label{eq:extchargesph}
\end{align}
is the radial component of the external charge density, with
\begin{align}
\alpha_{lm} & = \sqrt{\frac{2l+1}{4 \pi} \frac{(l-m)!}{(l+m)!}} , \label{eq:alphalm} \\
\Pi_{lm} (r) & = g_{lm}\bigg{(}\frac{\omega}{v} \sqrt{r^2 - b^2}\bigg{)} P_{l}^{m}\bigg{(}\sqrt{1 - \frac{b^2}{r^2}} \bigg{)}, \label{eq:Pilm} \\
g_{lm}(x) & = \begin{cases} \cos(x), & \textrm{ if } l + m \textrm{ is even}, \\
i \sin(x), & \textrm{ if } l + m \textrm{ is odd}, \label{eq:glm}
\end{cases}
\end{align}
$\Theta(x)$ the Heaviside step function, and $P_{l}^{m}(x)$ is the associated Legendre polynomial of degree $l$ and order $m$.

\subsection{Multipole expansion of the boundary value problem} \label{app:a.2}
We expand the induced charge density in terms of spherical harmonics:
\begin{align}
\varrho^{\textrm{ind}}(\textbf{r}) & = \sum_{l=0}^{\infty} \sum_{m=-l}^{l} \varrho^{\textrm{ind}}_{lm}(r) Y_{l}^{m} (\Omega), \label{eq:chargedensitygeneral}
\end{align}
where $\varrho^{\textrm{ind}}_{lm}(r)$ is its radial component.
The radial part of Helmholtz's equation [Eq.~\eqref{eq:Helmholtz}] reduces to the inhomogeneous spherical Bessel differential equation,
\begin{align}
\bigg{[} \frac{d^2}{dr^2} + \frac{2}{r} \frac{d}{dr} - \frac{l(l+1)}{r^2} + \mu^2 \bigg{]} \varrho^{\textrm{ind}}_{lm}(r) = \frac{\omega_{p}^{2}}{\beta^{2}} \varrho_{lm}^{\textrm{ext}}(r). \label{eq:bessel_differential}
\end{align}
To compute the boundary condition given by Eq.~\eqref{eq:BC_hydro}, we expand Eq.~\eqref{eq:potential_integral} in terms of spherical harmonics:
\begin{align}
\phi(\textbf{r}) = \sum_{l=0}^{\infty} \sum_{m=-l}^{l} Y_{l}^{m}(\Omega) \phi_{lm} (r), \label{eq:pot_sphharm}
\end{align}
with the radial component of the potentials given by
\begin{align}
\nonumber \phi_{lm}(r) = \; & \frac{4 \pi}{2 l + 1} \int_{0}^{\infty} dr' \, r'^2 \frac{r_{<}^{l}}{r_{>}^{l+1}} \varrho_{lm}(r'), \\
\nonumber = \; & \frac{4 \pi}{2 l + 1} \bigg{\{} \frac{1}{r^{l+1}} \int_{0}^{r} dr' \, r'^{l+2} \varrho_{lm}(r') \\
& + r^{l} \int_{r}^{\infty} \frac{dr'}{r'^{l-1}} \varrho_{lm}(r') \bigg{\}}. \label{eq:pot_radial}
\end{align}
Here, we make use of the expansion of the charge densities [Eqs.~\eqref{eq:chargedensityextgeneral} and \eqref{eq:chargedensitygeneral}] and the Coulomb kernel $1/\vert \textbf{r} - \textbf{r}' \vert$ in terms of spherical harmonics,
\begin{align}
\frac{1}{\vert \textbf{r} - \textbf{r}' \vert} & = \sum_{l=0}^{\infty} \sum_{m=-l}^{l} \frac{4 \pi}{2 l + 1} \frac{r_{<}^{l}}{r_{>}^{l+1}} Y_{l}^{m}(\Omega') Y_{l}^{m*}(\Omega),\label{eq:coulomb_sphharm}
\end{align}
where $r_{<}=\min(r,r')$ and $r_{>}=\max(r,r')$.
Introducing these expansions in Eq.~\eqref{eq:BC} for $r=a$ and applying the Leibniz integral rule \cite{Abramowitz_1972}, we write
\begin{align}
\nonumber \hspace{1em}&\hspace{-1em} \frac{\omega_p^2}{2l+1} \frac{l+1}{a^{l+2}} \int_{0}^{a} dr \, r^{l+2} \varrho^{\textrm{ind}}_{lm}(r) - \beta^2 \frac{\partial \varrho^{\textrm{ind}}_{lm}(r)}{\partial r} \bigg{\vert}_{r=a} = \\
\nonumber = & \; \frac{\omega_p^2}{2l+1} \bigg{\{} - \frac{l+1}{a^{l+2}} \int_{0}^{a} dr \, r^{l+2} \varrho_{lm}^{\textrm{ext}}(r) \\
& \hspace{4em}+ l \, a^{l-1} \int_{a}^{\infty} \frac{dr}{r^{l-1}} \varrho_{lm}^{\textrm{ext}}(r) \bigg{\}}, \label{eq:BCspherical}
\end{align}
which reduces the boundary condition to a purely radial condition for each multipole.
Equation \eqref{eq:BCspherical} stands for the general case of penetrating trajectories, which requires separating the external potential in Eq.~\eqref{eq:BC} into two contributions, one where the electron beam is outside the nanoparticle ($r>a$), and the other where it is inside the nanoparticle ($r<a$).
Notice that the Heaviside step function in Eq.~\eqref{eq:extchargesph} sets the condition $r>b$ in the integrals involving $\varrho^{\textrm{ext}}_{lm}(r)$.
Thus, the lower limit of the first integral of the right-hand side of Eq.~\eqref{eq:BCspherical} becomes $b$ instead of $0$ when introducing the explicit expression of $\varrho^{\textrm{ext}}_{lm}(r)$ with the Heaviside step function.
This constraint has to be taken into account for calculating the integrals or derived expressions that involve the explicit expression of $\varrho_{lm}^{\textrm{ext}}(r)$ (see also Appendix \ref{app:a.6}).

In addition to the hydrodynamic boundary condition, the induced charge density must also satisfy the \textit{charge conservation principle}; i.e., the net induced charge in the volume of the nanoparticle, $V$, has to be zero:
\begin{align}
    \int_{V} d\textbf{r} \, \varrho^{\textrm{ind}}(\textbf{r}) = 0. \label{eq:chargeneutrality}
\end{align}
Due to the symmetry of the $l>0$ terms of the spherical harmonics, their contribution to this integral is trivially zero; i.e., only the $\varrho_{00}(r)$ term has a nonzero contribution, leading to
\begin{align}
    \int_{0}^{a} dr \, r^{2} \, \varrho^{\textrm{ind}}_{00}(r)= 0,
\end{align}
and the boundary condition for the $l=0$ term as
\begin{align}
    \frac{\partial \varrho^{\textrm{ind}}_{00} (r)}{\partial r} \bigg{\vert}_{r=a} & = \frac{\omega_p^2}{a^2  \beta^2} \int_{0}^{a} dr \, r^{2} \, \varrho_{00}^{\textrm{ext}}(r) . \label{eq:BCspherical_l0}
\end{align}
This constraint does not hold for the $l>0$ terms, for which the boundary condition is given by Eq.~\eqref{eq:BCspherical}.

\subsection{General solution of the inhomogeneous boundary value problem}
The solution of the inhomogeneous boundary value problem is obtained in terms of a \textit{particular} solution $\varrho^{p}(\textbf{r})$ that satisfies the inhomogeneous Helmholtz equation,
\begin{align}
[\nabla^2 + \mu^2] \varrho^{p}(\textbf{r}) & = \frac{\omega_p^2}{\beta^2} \varrho^{\textrm{ext}}(\textbf{r}),
\end{align}
accompanied by a \textit{complementary} solution $\varrho^{c}(\textbf{r})$ (also referred to as \textit{homogeneous} solution) that satisfies the homogeneous Helmholtz equation,
\begin{align}
    [\nabla^2 + \mu^2] \varrho^{c}(\textbf{r}) & = 0. \label{eq:Helmholtz_complementary}
\end{align}
The particular solution $\varrho^{p}(\textbf{r})$ is calculated using the Green's function $\mathcal{G}(\textbf{r}, \textbf{r}')$ that satisfies
\begin{align}
[\nabla^2 + \mu^2] \mathcal{G}(\textbf{r}, \textbf{r}') = \delta(\textbf{r}, \textbf{r}'). \label{eq:HelmholtzGreen}
\end{align}
Therefore, the general solution is written as  \cite{Jackson_1999_book}
\begin{align}
    \varrho^{\textrm{ind}}(\textbf{r}) = \underbrace{\int d\textbf{r}' \mathcal{G}(\textbf{r}, \textbf{r}') \frac{\omega_{p}^{2}}{\beta^{2}} \varrho^{\textrm{ext}}(\textbf{r}')}_{\varrho^{p}(\textbf{r})} + \varrho^{c}(\textbf{r}). \label{eq:charge_density_Green}
\end{align}

Exploiting the spherical symmetry of the problem, $\mathcal{G}(\textbf{r}, \textbf{r}')$ is written in terms of spherical harmonics:
\begin{equation}
\mathcal{G}(\textbf{r}, \textbf{r}') = \sum^{\infty}_{l=0} \sum_{m=-l}^{l} Y^{m*}_{l}(\Omega ') \; Y^{m}_{l}(\Omega) G_{l}(r, r'), \label{eq:Green_function_sphharm}
\end{equation}
where its radial component $G_{l}(r,r')$ (also referred to as the reduced Green's function) satisfies the spherical Bessel differential equation,
\begin{equation}
\bigg{[} \frac{d^2}{dr^2} + \frac{2}{r} \frac{d}{dr} - \frac{l(l+1)}{r^2} + \mu^2 \bigg{]} G_{l}(r,r') = \frac{1}{r^2} \delta (r - r'). \label{eq:bessel_Green}
\end{equation}
Expanding the particular solution $\varrho^{p}(\textbf{r})$ in terms of spherical harmonics and taking into account Eqs.~\eqref{eq:charge_density_Green} and \eqref{eq:Green_function_sphharm}, the radial component of the particular solution $\varrho^{p}_{lm}(r)$ reduces to
\begin{align}
\varrho^{p}_{lm}(r) = \frac{\omega_p^2}{\beta^2} \int_{0}^{a} dr' \, r'^2 G_{l}(r,r') \varrho_{lm}^{\textrm{ext}}(r'). \label{eq:particular_sphharm}
\end{align}
On the other hand, the radial component of the complementary solution $\varrho^{c}_{lm}(r)$ must satisfy the homogeneous spherical Bessel differential equation [Eq.~\eqref{eq:bessel_differential} with $\varrho^{\textrm{ext}}_{lm}(r)=0$].
Imposing regularity at the solution, the complementary solution is written in terms of the spherical Bessel functions of the first kind $j_{l}(x)$ [the spherical Bessel functions of the second kind $y_{l}(x)$ diverge at the origin]:
\begin{align}
    \varrho^{c}_{lm}(r) = d_{lm} \, j_{l}(\mu r), \label{eq:complementary_sphharm}
\end{align}
with $d_{lm}$ a coefficient to be determined by the boundary conditions.
Therefore, the radial part of the induced charge density $\varrho_{lm}$ is written as
\begin{align}
\varrho^{\textrm{ind}}_{lm}(r) = & \; \varrho^{p}_{lm}(r) + \varrho^{c}_{lm}(r). \label{eq:charge_density_Green_sphharm}
\end{align}

\subsection{Evaluation of the boundary conditions}
Although Green's functions must satisfy the homogeneous boundary condition, there is again a constraint for the \hbox{$l=0$} term.
By integrating Eq.~\eqref{eq:HelmholtzGreen} over a sphere of radius $r=a$ and then applying the divergence theorem, we write \cite{Kim1993}
\begin{align}
a^2 \frac{\partial G_{0}(r,r')}{\partial r} \bigg{\vert}_{r=a} & = 1 - \mu^2 \int_{0}^{a} dr \, r^2 G_{0}(r,r').
\end{align}
From a physical perspective, the Green's function may be interpreted as the charge density at position $r$ induced by a point charge at $r'$; thus, the integral in the right-hand side has to be zero for any $r'$, as it is the total induced charge.
Therefore,
\begin{align}
\int_{0}^{a} dr \, r^2 G_{0}(r,r') = 0.
\end{align}
This condition provides the following boundary condition:
\begin{align}
\frac{\partial G_{0}(r,r')}{\partial r} \bigg{\vert}_{r=a} = \frac{1}{a^2}. \label{eq:BC_green_l0}
\end{align}
Notice that this condition ensures charge conservation for the particular solution, when integrating over all $\textbf{r}'$ positions of the external source:
\begin{align}
    \int_{0}^{a} \varrho_{00}^{p}(r) r^2 dr & = 0.
\end{align}
As a consequence, the $l=0$ term of the complementary solution must satisfy
\begin{align}
    \int_{0}^{a} \varrho_{00}^{c}(r) r^2 dr = 0.
\end{align}
By substituting the explicit expression of $\varrho_{00}(r)$ [Eq. \eqref{eq:complementary_sphharm}] into the integral, and applying the properties of the spherical Bessel functions \cite{Abramowitz_1972}, we obtain the condition
\begin{align}
    d_{00}\int_{0}^{a} j_{0}(\mu r) r^2 dr = d_{00} \; j_{1}(\mu a)=0 ,\label{eq:condition_complementary_l0}
\end{align}
which has to be satisfied for any $\omega$.
Therefore, this imposes \hbox{$d_{00} = 0$} and \hbox{$\varrho^{c}_{00}(r)= 0$}.
Consequently, the boundary condition of the $l=0$ term of the particular solution reduces to
\begin{align}
    \frac{\partial \varrho_{00}^{p} (r)}{\partial r} \bigg{\vert}_{r=a} & = \frac{\omega_p^2}{a^2  \beta^2} \int_{0}^{a} \varrho_{00}^{\textrm{ext}}(r) r^2 dr.
\end{align}

On the other hand, for the $l>0$ terms of the Green's function there is not a similar constraint.
For these terms, we impose homogeneous boundary conditions at the surface $r=a$:
\begin{align}
\nonumber \frac{l+1}{2l+1} \frac{\omega_p^2}{a^{l+2}} \int_{0}^{a} dr \, r^{l+2} G_{l}(r, r') & \\
\hspace{6em} & \hspace{-6em}  - \beta^2 \frac{\partial G_{l}(r,r')}{\partial r} \bigg{\vert}_{r=a} = 0. \label{eq:BC_green_l}
\end{align}
This condition sets the following homogeneous boundary conditions for the $l>0$ terms of the particular solution, $\varrho^{p}_{lm}(r)$:
\begin{align}
\nonumber \frac{\omega_p^2}{2l+1} \frac{l+1}{a^{l+2}} \int_{0}^{a} dr \, r^{l+2} \varrho_{lm}^{p}(r) & \\
\hspace{6em} & \hspace{-6em} - \beta^2 \frac{\partial \varrho_{lm}^{p}(r)}{\partial r} \bigg{\vert}_{r=a} = 0.
\end{align}
Therefore, by construction the $l>0$ terms of $\varrho^{c}_{lm}(r)$ must satisfy the following inhomogeneous boundary condition:
\begin{align}
\nonumber \hspace{2em}&\hspace{-2em} \frac{\omega_p^2}{2l+1} \frac{l+1}{a^{l+2}} \int_{0}^{a} dr \, r^{l+2} \varrho^{c}_{lm}(r) - \beta^2 \frac{\partial \varrho^{c}_{lm}(r)}{\partial r} \bigg{\vert}_{r=a} = \\
\nonumber = & \; \frac{\omega_p^2}{2l+1} \bigg{\{} - \frac{l+1}{a^{l+2}} \int_{0}^{a} dr \, r^{l+2} \varrho_{lm}^{\textrm{ext}}(r) \\
& + l \, a^{l-1} \int_{a}^{\infty} \frac{dr}{r^{l-1}} \varrho_{lm}^{\textrm{ext}}(r) \bigg{\}}. \label{eq:BC_complemetary}
\end{align}

\subsection{The Green's function}
The Green's function that satisfies Eq.~\eqref{eq:bessel_Green} in the sphere has the form:
\begin{align}
G_{l}(r,r') =
\begin{cases}
    a_{l}(r') \; j_{l}(\mu r), \hspace{1em} & 0 < r < r' < a, \\
    b_{l}(r') \; j_{l}(\mu r) \\
    + \, c_{l}(r') \; y_{l}(\mu r) , \hspace{1em} & 0 < r' < r < a, \\
    0, \hspace{1em} & r , r' > a,
\end{cases} \label{eq:Green_function}
\end{align}
where we have assumed regularity at $r=0$.
Here, $j_{l}(x)$ and $y_{l}(x)$ are the spherical Bessel functions of the first and second kind, respectively.
The coefficients $a_{l}$, $b_{l}$ and $c_{l}$ are determined from the boundary condition [Eq.~\eqref{eq:BC_green_l0} for $l=0$ and Eq.~\eqref{eq:BC_green_l} for $l>0$], the continuity of the Green's functions at \hbox{$r=r'$}, and the discontinuity of their derivative at \hbox{$r=r'$}.
After rather evolved but elementary calculations based on the properties of Bessel functions, as their cross products and together with the following integrals, with $f_{l}$ standing for $j_{l}$ or $y_{l}$,
\begin{align}
\int dx \; x^{l+2} f_l(x) = & \; x^{l+2} f_{l+1}(x), \label{eq:sphBessel_int1} \\ 
\int dx \; \frac{f_l(x)}{x^{l-1}}  = & - \frac{f_{l-1} (x)}{x^{l-1} } ,\label{eq:sphBessel_int2}
\end{align}
we obtain for $l=0$,
\begin{subequations}
\begin{align}
\nonumber a_{0}(r') = \; &  \mu y_{0}(\mu r') - \frac{y_{1}(\mu a)}{j_{1}(\mu a)} \mu j_{0}(\mu r')\\
& - \frac{1}{a^2 \mu j_{1}(\mu a)}, \label{eq:coefficient_a_Green_l0}\\
b_{0}(r') = & - \frac{y_{1}(\mu a)}{j_{1}(\mu a)} \mu j_{0}(\mu r') - \frac{1}{a^2 \mu j_{1}(\mu a)},  \label{eq:coefficient_b_Green_l0}\\
c_{0}(r') = & \mu j_0(\mu r'), \label{eq:coefficient_c_Green_l0}
\end{align}
\end{subequations}
and for $l>0$,
\begin{subequations}
\begin{align}
\nonumber a_{l}(r') = & \mu y_l(\mu r') - \frac{N_{l}}{M_{l}} \mu j_l(\mu r') \\
& - \frac{\omega_p^2}{\mu \, M_{l}}\frac{l+1}{2l+1} \frac{r'^{l}}{a^{l+2}}, \label{eq:coefficient_a_Green_l}\\
b_{l}(r') = & - \frac{N_{l}}{M_{l}} \mu j_l(\mu r') - \frac{\omega_p^2}{\mu \, M_{l}}\frac{l+1}{2l+1} \frac{r'^{l}}{a^{l+2}}, \label{eq:coefficient_b_Green_l}\\
c_{l}(r') = & \mu j_l(\mu r') \label{eq:coefficient_c_Green_l},
\end{align}
\end{subequations}
where $N_{l}$ and $M_{l}$ are given by Eqs.~\eqref{eq:Nl} and \eqref{eq:Ml}.

\subsection{Induced charge density} \label{app:a.6}

The particular solution is obtained through Eq.~\eqref{eq:particular_sphharm} using the Green's function $G_{l}(r,r')$ given by Eq.~\eqref{eq:Green_function} and the external charge density $\varrho^{\textrm{ext}}_{lm}(r)$ [Eq.~\eqref{eq:extchargesph}].
Notice that to compute the integral over the path of the electron beam in Eq.~\eqref{eq:particular_sphharm}, we have to take into account that the Green's function depends on whether $r>r'$ or $r<r'$, where $r'$ corresponds to the position of the beam.
Indeed, the presence of the Heaviside step function in $\varrho^{\textrm{ext}}_{lm}(r')$ (see Appendix \ref{app:a.1}) sets the condition $r'>b$ in the integrals, and we have regions where $r'>b>r$.
Therefore, one has to divide the integral into three different intervals, $r<b<r'$, $b<r'<r$, and $b<r<r'$, which leads to
\begin{widetext}
\begin{alignat}{1}
\nonumber \varrho^{p}_{lm}(r) = \frac{-2\omega_p^2 \alpha_{lm}}{v \beta^2} \bigg{\{} & \Theta \bigg{(} 1 -\frac{r}{b} \bigg{)} \underbrace{j_{l}(\mu r) \int_{b}^{a} dr' \, r'^2 a_{l}(r') \varrho_{lm}^{\textrm{ext}}(r')}_{r<b<r'<a} + \Theta \bigg{(}\frac{r}{b} - 1 \bigg{)} \bigg{[} \underbrace{j_{l}(\mu r) \int_{r}^{a} dr' \, r'^2 a_{l}(r') \varrho_{lm}^{\textrm{ext}}(r')}_{b<r<r'<a} \\
+ \, & \underbrace{j_{l}(\mu r) \int_{b}^{r} dr' \, r'^2 b_{l}(r') \varrho_{lm}^{\textrm{ext}}(r') + y_{l}(\mu r) \int_{b}^{r} dr' \, r'^2 c_{l}(r') \varrho_{lm}^{\textrm{ext}}(r')}_{b<r'<r<a} \bigg{]} \bigg{\}}.
\end{alignat}
\end{widetext}
By substituting the coefficients $a_{l}(r')$, $b_{l}(r')$, and $c_{l}(r')$, given by Eqs.~\eqref{eq:coefficient_a_Green_l0}-\eqref{eq:coefficient_c_Green_l0} for $l=0$ and Eqs.~\eqref{eq:coefficient_a_Green_l}-\eqref{eq:coefficient_c_Green_l} for $l>0$, and the radial component of the external charge density $\varrho^{\textrm{ext}}_{lm}(r)$ [Eq.~\eqref{eq:extchargesph}], we obtain the particular solution $\varrho^{p}_{lm}(r)$:
\begin{alignat}{2}
    \nonumber \varrho_{lm}^{p}(r) = \frac{2 \omega_p^2 \, \alpha_{lm}}{v} \bigg{\{} & \bigg{[} && \bigg{(} \delta(l) +\frac{\omega^2_p}{\mu^2 \beta^2} \bigg{)} A_{lm} + \delta(l) \, B_{lm} \\
    \nonumber & && + C_{lm} \bigg{]}\, j_{l}(\mu r) \\
    & - && \, \Theta \Big{(} \frac{r}{b} - 1 \Big{)} \mathcal{D}_{lm}(r) \bigg{\}}, \label{eq:charge_particular_spherical}
\end{alignat}
where $\delta(x)$ is the Dirac delta function, the coefficients $A_{lm}$, $B_{lm}$, and $C_{lm}$ correspond to
\begin{subequations}
\begin{align}
    A_{lm} = & \; \frac{\mu}{a M_{l}} \frac{l+1}{2l+1} \mathcal{I}_{lm}(b,a), \label{eq:coeff_A} \\
    B_{lm} = & \; - \frac{\mu}{a M_{l}} \frac{l}{2l+1} \mathcal{O}_{lm}(a,\infty), \label{eq:coeff_B} \\
    C_{lm} = & \; \frac{1}{\beta^2 M_{l}} [N_{l} \mathcal{J}_{lm}(b,a) - M_{l} \mathcal{Y}_{lm}(b,a)], \label{eq:coeff_C}
\end{align}
the function $\mathcal{D}_{lm}$ is defined as
\begin{align}
    \nonumber \mathcal{D}_{lm}(r) = \frac{1}{\beta^2} \bigg{[} & y_{l}(\mu r) \mathcal{J}_{lm}(b,r) \\
    & \, - j_{l}(\mu r) \mathcal{Y}_{lm}(b,r) \bigg{]}, \label{eq:coeff_D}
\end{align}
\end{subequations}
and $\mathcal{I}_{lm}$, $\mathcal{O}_{lm}$, $\mathcal{J}_{lm}$, and $\mathcal{Y}_{lm}$ are defined as
\begin{subequations}
\begin{align}
    \mathcal{I}_{lm}(r_{1},r_{2}) & \equiv \int_{r_{1}}^{r_{2}} dr \, \frac{r^{l}}{a^{l+1}} \frac{\Pi_{lm}(r)}{\sqrt{1 - b^2 / r^2}} , \label{eq:Ilm}\\
    \mathcal{O}_{lm}(r_{1},r_{2}) & \equiv \int_{r_{1}}^{r_{2}} dr \frac{a^{l}}{r^{l+1}} \frac{\Pi_{lm}(r)}{\sqrt{1 - b^2 / r^2}} , \label{eq:Olm} \\
    \mathcal{J}_{lm} (r_{1}, r_{2}) & \equiv \int_{r_{1}}^{r_{2}} dr \, \mu \, j_{l}(\mu r) \frac{\Pi_{lm} (r)}{\sqrt{1 - b^2 / r^2}}, \label{eq:Jlm} \\
    \mathcal{Y}_{lm} (r_{1}, r_{2}) & \equiv \int_{r_{1}}^{r_{2}} dr \, \mu \, y_l(\mu r) \frac{\Pi_{lm} (r)}{\sqrt{1 - b^2 / r^2}}. \label{eq:Ylm}
\end{align}
\end{subequations}

Additionally, we calculate the complementary solution by solving the inhomogeneous boundary condition [Eq. \eqref{eq:BC_complemetary}] for $\varrho^{c}_{lm}(r)$ [Eq.~\eqref{eq:complementary_sphharm}] with $\varrho^{\textrm{ext}}_{lm}(r)$ [Eq.~\eqref{eq:extchargesph}] to obtain the $d_{lm}$ coefficients for $l>0$ [notice that $d_{00} = 0$, as imposed by Eq.~\eqref{eq:condition_complementary_l0}]:
\begin{align}
    \varrho^{c}_{lm} = \frac{2 \omega_{p}^{2} \alpha_{lm}}{v} [A_{lm} + B_{lm}] \, j_{l}(\mu r) && (l>0), \label{eq:chargedensityhomogeneous}
\end{align}
where we have made use of Eq.~\eqref{eq:sphBessel_int1}.

Adding the complementary solution to the particular solution [Eq.~\eqref{eq:charge_particular_spherical}] yields the following expression for the induced charge density:
\begin{alignat}{2}
    \nonumber \varrho^{\textrm{ind}}_{lm}(r) = \frac{2 \omega_p^2 \, \alpha_{lm}}{v} \bigg{\{} & \bigg{[} && \bigg{(} 1 +\frac{\omega^2_p}{\mu^2 \beta^2} \bigg{)} A_{lm} +  B_{lm} \\
    \nonumber & && + C_{lm} \bigg{]}\, j_{l}(\mu r) \\
    & - && \, \Theta \Big{(} \frac{r}{b} - 1 \Big{)} \mathcal{D}_{lm}(r) \bigg{\}}. \label{eq:chargedensitytotal_appendix}
\end{alignat}
This corresponds to Eq.~\eqref{eq:chargedensitytotal}, where the coefficients $A_{lm}$, $B_{lm}$, and $C_{lm}$, and the function $\mathcal{D}_{lm}$ are given by Eqs.~\eqref{eq:coeff_A}-\eqref{eq:coeff_D}.

\begin{widetext}
\section{The induced potential} \label{app:b}
The induced potential can be directly calculated via Coulomb's potential using Eq.~\eqref{eq:potential_integral}.
Using its decomposition in spherical harmonics [see Eq.~\eqref{eq:indpotsphharm}], with the radial component given by Eq.~\eqref{eq:indporadial}, yields the radial component of the induced potential outside the nanoparticle $\phi^{\textrm{ind},o}_{lm}(r)$:
\begin{align}
    \phi^{\textrm{ind},o}_{lm}(r) = \frac{4 \pi}{2l+1} \frac{1}{r^{l+1}} \int_{0}^{a} dr' r'^{l+2} \varrho^{\textrm{ind}}_{lm}(r').
\end{align}
After substituting $\varrho^{\textrm{ind}}_{lm}(r)$ given by Eq.~\eqref{eq:chargedensitytotal_appendix} we obtain
\begin{align}
    \phi^{\textrm{ind},o}_{lm}(r) = \frac{L_{lm}} {r^{l+1}} \bigg{\{} E_{lm}\int_{0}^{a} dr' r'^{l+2} j_{l}(\mu r') - \int_{b}^{a} dr' \, r'^{l+2} \mathcal{D}_{lm}(r') \bigg{\}}, \label{eq:indpotout_appendix_expansion}
\end{align}
with
\begin{align}
    L_{lm} = & \; \frac{8 \pi \omega_{p}^{2} \alpha_{lm}}{v(2l+1)} , \\
    E_{lm} = & \; \bigg{[} \bigg{(} 1 +\frac{\omega^2_p}{\mu^2 \beta^2} \bigg{)} A_{lm} +  B_{lm} + C_{lm} \bigg{]}.
\end{align}
The first integral in Eq.~\eqref{eq:indpotout_appendix_expansion} is readily calculated with Eq.~\eqref{eq:sphBessel_int1}.
The second one involves a double integral, which is reduced to
\begin{align}
    \int_{b}^{a} dr' \, r'^{l+2} \mathcal{D}_{lm}(r') = \frac{1}{\beta^{2}} \int_{b}^{a} dr' \, r'^{l+2} \Big{[} y_{l}(\mu r') \mathcal{J}_{lm}(b,r') - j_{l}(\mu r') \mathcal{Y}_{lm}(b,r') \Big{]},
\end{align}
with $\mathcal{J}_{lm}$ and $\mathcal{Y}_{lm}$ given by Eqs.~\eqref{eq:Jlm} and \eqref{eq:Ylm}.
More explicitly, we write
\begin{align}
    \nonumber \int_{0}^{a}dr' \, r'^{l+2} y_{l}(\mu r') \mathcal{J}_{lm}(\mu r') & = \int_{b}^{a}dr' \, r'^{l+2} y_{l}(\mu r') \int_{b}^{r'} dr''\mu \, j_{l}(\mu r'') \frac{\Pi_{lm} (r'')}{\sqrt{1 - b^2 / r''^2}} \\
    & = \int_{b}^{a}dr' \, \mu \, j_{l}(\mu r') \frac{\Pi_{lm} (r')}{\sqrt{1 - b^2 / r'^2}} \int_{r'}^{a} dr'' r''^{l+2} y_{l}(\mu r''),
\end{align}
so that, by applying Eq.~\eqref{eq:sphBessel_int1}, the double integral reduces to a single one.
After some straightforward but involved calculations, using cross products, identities, and recurrence relations of the spherical Bessel functions, we obtain
\begin{align}
\phi_{lm}^{\textrm{ind},o}(r) & = \Phi_{lm}^{o} \bigg{(} \frac{a}{r} \bigg{)}^{l+1}, \label{eq:indpotout_appendix}
\end{align}
where
\begin{align}
    \Phi_{lm}^{o} = \frac{L_{lm}}{M_{l}} \bigg{\{} \frac{j_{l-1}(\mu a)}{2l+1} \mathcal{I}_{lm}(b,a) - \frac{j_{l+1}(\mu a)}{2l+1} \, \mathcal{O}_{lm}(a,\infty) - \frac{1}{\mu^2 a^2} \mathcal{J}_{lm}(b,a) \bigg{\}}.
\end{align}

We proceed in a similar way to obtain the radial component of the induced potential inside the nanoparticle, $\phi^{\textrm{ind},i}_{lm}(r)$, starting from Eq.~\eqref{eq:indporadial}:
\begin{align}
    \phi^{\textrm{ind},i}_{lm}(r) = \frac{4 \pi}{2l+1} \bigg{[} & \frac{1}{r^{l+1}} \int_{0}^{r} dr' r'^{l+2} \varrho^{\textrm{ind}}_{lm}(r') + r^{l} \int_{r}^{a} \frac{dr'}{r'^{l-1}}\varrho^{\textrm{ind}}_{lm}(r') \bigg{]}.
\end{align}
Using $\varrho^{\textrm{ind}}_{lm}(r)$ given by Eq.~\eqref{eq:chargedensitytotal_appendix} we obtain
\begin{align}
    \nonumber \phi^{\textrm{ind},i}_{lm}(r) = & \; L_{lm}  \bigg{\{} E_{lm} \bigg{[} \frac{1}{r^{l+1}} \int_{0}^{r} dr' r'^{l+2} j_{l}(\mu r') + r^{l} \int_{r}^{a} \frac{dr'}{r'^{l-1}} j_{l}(\mu r') \bigg{]} +  \Theta\bigg{(} 1 - \frac{r}{b}\bigg{)} r^{l} \int_{b}^{a} \frac{dr'}{r'^{l-1}} \mathcal{D}_{lm}(r') \\
    & + \Theta\bigg{(} \frac{r}{b} - 1 \bigg{)} \bigg{[} \frac{1}{r^{l+1}} \int_{b}^{r} dr' \, r'^{l+2} \mathcal{D}_{lm}(r') + r^{l} \int_{r}^{a} \frac{dr'}{r'^{l-1}} \mathcal{D}_{lm}(r') \bigg{]} \bigg{\}}. \label{eq:indpotin_appendix_expansion}
\end{align}
Finally, we evaluate the double integrals as explained for $\phi^{\textrm{ind},o}_{lm}(r)$ and, after some algebra, we obtain
\begin{align}
\nonumber \phi^{\textrm{ind},i}_{lm}(r) = & \; \Phi_{lm}^{i,1} \, j_{l}(\mu r) + \Phi_{lm}^{i,2} \bigg{(} \frac{r}{a} \bigg{)} ^{l} - \Theta \Big{(} \frac{r}{b} - 1 \Big{)} L_{lm} \frac{2l+1}{\mu^2 \beta^2} \Bigg{[} y_{l}(\mu r) \mathcal{J}_{lm}(b,r) - j_{l}(\mu r) \mathcal{Y}_{lm}(b,r) \\
& + \frac{\mathcal{I}_{lm}(b,r)}{2l+1} \bigg{(} \frac{a}{r} \bigg{)}^{l+1} - \frac{\mathcal{O}_{lm}(b,r)}{2l+1} \bigg{(} \frac{r}{a} \bigg{)}^{l} \Bigg{]}, \label{eq:indpotin_appendix}
\end{align}
where
\begin{align}
    \Phi_{lm}^{i,1} = \frac{(2l+1)}{\mu^{2}} L_{lm} E_{lm} ,
\end{align}
\begin{align}
    \nonumber \Phi_{lm}^{i,2} = & \; L_{lm} \bigg{[} - \frac{(l+1) j_{l-1}(\mu a)}{(2l+1)M_{l}} \bigg{(} 1 + \frac{\omega^2_p}{\mu^2 \beta^2} \bigg{)} \mathcal{I}_{lm}(b,a) + \frac{l \, j_{l-1}(\mu a)}{(2l+1) M_{l}} \mathcal{O}_{lm}(a,\infty) - \frac{1}{\mu^2 \beta^2} \mathcal{O}_{lm}(b,a) \\
    & + \frac{l+1}{\mu^2 a^2 M_{l}} \bigg{(} 1 + \frac{\omega^2_p}{\mu^2 \beta^2} \bigg{)} \mathcal{J}_{lm}(b,a) \bigg{]} .
\end{align}

\section{Calculation of the electron energy loss probability} \label{app:c}
The EEL probability is calculated by separating the path of the electron inside and outside the nanoparticle in Eq.~\eqref{eq:eelsspherical}:
\begin{align}
\Gamma_{\textrm{EELS}} = & \; \frac{2}{\pi v} \sum_{l=0}^{\infty} \sum_{m=-l}^{l} \alpha_{lm} (-1)^{l+m+1} \textrm{Im} \bigg{\{} \int^{\infty}_{z_{a}} dz \, \phi^{\textrm{ind,o}}_{lm}(r) P_{l}^{m} \bigg{(} \frac{z}{r} \bigg{)} g_{lm}\bigg{(}\frac{\omega z}{v} \bigg{)} + \int_{0}^{z_{a}} dz \, \phi^{\textrm{ind,i}}_{lm}(r) P_{l}^{m} \bigg{(} \frac{z}{r} \bigg{)} g_{lm}\bigg{(}\frac{\omega z}{v} \bigg{)} \bigg{\}}, \label{eq:eelsspherical_in_out}
\end{align}
where the limit of integration $z_{a} = \sqrt{a^{2} - b^{2}}$.
By making a change of variable from $z$ to $r=\sqrt{z^{2}+b^{2}}$, the integrals are then written as
\begin{align}
    I_{lm}^{o} = \int_{a}^{\infty} dr \, \phi^{\textrm{ind,o}}_{lm}(r) \frac{\Pi_{lm}(r)}{\sqrt{1-b^{2}/r^{2}}},
\end{align}
\begin{align}
    I_{lm}^{i} = \int_{b}^{a} dr \, \phi^{\textrm{ind,i}}_{lm}(r) \frac{\Pi_{lm}(r)}{\sqrt{1-b^{2}/r^{2}}} .
\end{align}

Further introduction of the explicit expressions of $\phi^{\textrm{ind,o}}_{lm}(r)$ and $\phi^{\textrm{ind,i}}_{lm}(r)$ given by Eqs.~\eqref{eq:indpotout} and \eqref{eq:indpotin} leads to
\begin{align}
    I_{lm}^{o} & = a\, \Phi_{lm}^{o} \mathcal{O}_{lm}(a,\infty),
\end{align}
and
\begin{align}
    \nonumber I_{lm}^{i} = & \; \frac{\Phi_{lm}^{i,1}}{\mu} \mathcal{J}_{lm}(b,a) + a \, \Phi_{lm}^{i,2} \mathcal{I}_{lm}(b,a) - \frac{L_{lm}}{\mu^{2} \beta^{2}} \bigg{\{} (2l+1) \int_{b}^{a} dr \, \Big{[} y_{l}(\mu r) \mathcal{J}_{lm}(b,r) - j_{l}(\mu r) \mathcal{Y}_{lm}(b,r) \Big{]} \frac{\Pi_{lm}(r)}{\sqrt{1-b^{2}/r^{2}}} \\
    & + \int_{b}^{a} dr \, \bigg{[}\mathcal{I}_{lm}(b,r) \bigg{(} \frac{a}{r} \bigg{)}^{l+1} - \mathcal{O}_{lm}(b,r) \bigg{(} \frac{r}{a} \bigg{)}^{l} \bigg{]} \frac{\Pi_{lm}(r)}{\sqrt{1-b^{2}/r^{2}}} \bigg{\}} .
\end{align}
We notice that the number of required double integrals in $I_{lm}^{i}$ is reduced by considering
\begin{align}
    \mathcal{J}_{lm}(b,a) \mathcal{Y}_{lm}(b,a) = \mu \int_{b}^{a} & dr \, j_{l}(\mu r) \frac{\Pi_{lm}(r)}{\sqrt{1-b^{2}/r^{2}}} [\mathcal{Y}_{lm}(b,r) + \mathcal{Y}_{lm}(r,a)]
\end{align}
and
\begin{align}
    a \, \mathcal{I}_{lm}(b,a) \mathcal{O}_{lm}(b,a) = \int_{b}^{a} & dr \, \bigg{(} \frac{r}{a} \bigg{)} ^{l} \frac{\Pi_{lm}(r)}{\sqrt{1-b^{2}/r^{2}}} [\mathcal{O}_{lm}(b,r) + \mathcal{O}_{lm}(r,a)],
\end{align}
where we change the order of integration for the integral involving $\mathcal{Y}_{lm}(r,a)$ and $\mathcal{O}_{lm}(r,a)$, respectively, to obtain
\begin{align}
    \mathcal{J}_{lm}(b,a) \mathcal{Y}_{lm}(b,a) = \mu \int_{b}^{a} & dr \, \frac{\Pi_{lm}(r)}{\sqrt{1-b^{2}/r^{2}}} [j_{l}(\mu r) \mathcal{Y}_{lm}(b,r) + y_{l}(\mu r) \mathcal{J}_{lm}(b,r)],
\end{align}
and,
\begin{align}
    a \, \mathcal{I}_{lm}(b,a) \mathcal{O}_{lm}(b,a) = \int_{b}^{a} & dr \, \frac{\Pi_{lm}(r)}{\sqrt{1-b^{2}/r^{2}}} \bigg{[} \bigg{(}\frac{r}{a} \bigg{)} ^{l} \mathcal{O}_{lm}(b,r) + \bigg{(} \frac{a}{r} \bigg{)} ^{l+1} \mathcal{I}_{lm}(b,r) \bigg{]}.
\end{align}

Taking into account these relations, $I_{lm}^{i}$ is
\begin{align}
    \nonumber I_{lm}^{i} = & \, \frac{\Phi_{lm}^{i,1}}{\mu} \mathcal{J}_{lm}(b,a) +a \, \Phi_{lm}^{i,2} \mathcal{I}_{lm}(b,a) - \frac{L_{lm}}{\mu^{2} \beta^{2}} \bigg{\{} \frac{(2l+1)}{\mu} \bigg{[} 2 \, \mathcal{H}_{lm} - \mathcal{J}_{lm}(b,a) \mathcal{Y}_{lm}(b,a)\bigg{]} \\
    & + 2 \mathcal{F}_{lm} - a \, \mathcal{I}_{lm}(b,a) \mathcal{O}_{lm}(b,a)\bigg{\}},
\end{align}
where, we have introduced,
\begin{align}
    \mathcal{F}_{lm} = & \int_{0}^{\sqrt{a^2-b^2}} dz \frac{1}{r^{l+1}} P_{l}^{m}\bigg{(}\frac{z}{r}\bigg{)} g_{lm}\bigg{(}\frac{\omega z}{v}\bigg{)} \int_{0}^{z} dz' r'^{l} P_{l}^{m}\bigg{(}\frac{z'}{r'}\bigg{)} g_{lm}\bigg{(}\frac{\omega z'}{v}\bigg{)} \label{eq:Flm}
\end{align}
and
\begin{align}
    \mathcal{H}_{lm} = & \int_{0}^{\sqrt{a^2-b^2}} dz \, \mu \, y_{l}(\mu r) P_{l}^{m}\bigg{(}\frac{z}{r}\bigg{)} g_{lm}\bigg{(}\frac{\omega z}{v}\bigg{)}\int_{0}^{z} dz' \, \mu \, j_{l}(\mu r') P_{l}^{m}\bigg{(}\frac{z'}{r'}\bigg{)} g_{lm}\bigg{(}\frac{\omega z'}{v}\bigg{)}. \label{eq:Hlm}
\end{align}
Finally, by introducing the explicit expressions of $I_{lm}^{o}$ and $I_{lm}^{i}$ in Eq.~\eqref{eq:eelsspherical_in_out}, and taking into account the following property of the associated Legendre polynomials:
\begin{align}
P_{l}^{-m} (x) = (-1)^{m} \frac{(l-m)!}{(l+m)!} P_{l}^{m}(x),
\end{align}
we obtain after some algebra the main result of the EEL probability given by Eqs.~\eqref{eq:EELprobabilitygen} and \eqref{eq:Gammabulk}-\eqref{eq:Gammaext}.
\end{widetext}

%



\end{document}